\documentclass[aps,prd,preprintnumbers,showpacs]{revtex4}
\usepackage{graphicx}

\begin{document}

%
%

\eprint{Nisho-2-2024}
\title{Axion Dark Matter and Plateau-Plateau Transition in Quantum Hall Effect}
\author{Aiichi Iwazaki}
\affiliation{International Economics and Politics, Nishogakusha University,\\ 
6-16 3-bantyo Chiyoda Tokyo 102-8336, Japan }   
\date{Nov.8, 2024}
\begin{abstract}
Axion dark matter inevitably generates electromagnetic radiation in quantum Hall effect experiments that use strong magnetic fields. Although these emissions are very weak, we have shown using a QCD axion model that they influence the plateau-plateau transition at low temperatures (below $100$ mK) in a system with a large surface area (greater than $10^{-3}\rm cm^2$) of two-dimensional electrons. By analyzing previous experiments that show saturation of the transition width $\Delta B$ as temperature and microwave frequency change, we provide evidence for the presence of axions. Notably, in most experiments without axion effects, the saturation frequency $f_s(T)$ is less than $1$ GHz at temperatures of $100$ mK or higher and for system sizes of 
$10^{-3}\rm cm^2$ or smaller. Additionally, the frequency $f_s(T)$ decreases with decreasing temperature or increasing system size.
However, there are experiments that show a saturation frequency $f_s(T)\simeq 2.4$GHz despite a low temperature of 35 mK and a large surface area of $6.6\times 10^{-3}\rm cm^2$ for the Hall bar. This identical frequency of approximately $2.4$ GHz has also been observed in different plateau transitions and in Hall bars of varying sizes. These unexpected results are caused by axion microwaves. The saturation frequency $f_s=m_a/2\pi$ of $\simeq 2.4$ GHz implies an axion mass of $\simeq 10^{-5}$eV. 
By comparing the axion effect with thermal effect on the width $\Delta B$,
we have shown the dominance of the axion effect over thermal effect at low temperature less than $50$mK.
The dominance of the axion effect is attributed to significant absorption of axion energy, which is proportional to the square of the number of electrons involved.


%

\end{abstract}
\hspace*{0.3cm}

\hspace*{1cm}

\maketitle

\section{introduction}
Axions are currently considered one of the most promising candidates for dark matter in the Universe, and their discovery could provide key insights into new physics beyond the Standard Model of particle physics. Although several axion models, including axion-like particles (ALPs), exist with masses spanning tens of orders of magnitude, this paper focuses specifically on the QCD axion. The QCD axion model not only addresses the strong CP problem in QCD, but it also remains the most realistic among the various axion models under consideration. As a pseudo-Goldstone boson arising from the Peccei-Quinn symmetry, the QCD axion naturally resolves the strong CP problem \cite{axion1, axion2, axion3}. The mass of the QCD axion is constrained within the range
$m_a=10^{-6}\mbox{eV} \sim10^{-3}$eV \cite{Wil, Wil1, Wil2}, corresponding to electromagnetic radiation in the microwave band with frequencies from $1\rm GHz \sim 10^3$GHz, produced through axion-photon conversion.

\vspace{0.1cm}
Despite numerous experimental efforts\cite{admx,carrack,haystac,abracadabra,organ,madmax,brass,cast,sumico,iwazaki01}, 
many approaches either lack the sensitivity required for axion detection or are restricted to a narrow mass range, even when high sensitivity is achieved.
Most experiments rely on a strong magnetic field to stimulate electromagnetic radiation\cite{sikivie} from axions. Notably, strong magnetic fields are also employed in quantum Hall effect (QHE) experiments\cite{von,girvin}.

In previous researches\cite{iwa,iwa2}, we proposed detecting axions through the integer quantum Hall effect. 
In QHE, two-dimensional electrons absorb microwaves emitted by axions. This absorption causes electrons in localized states to transition to extended states, thereby enhancing the Hall conductivity. As a result, the axion's presence can be inferred from an increase in Hall conductivity.

By closely examining plateau-to-plateau transitions observed in past experiments, we can identify possible axion signatures in these transitions. In fact, our analysis suggests that existing experimental results may already indicate the presence of axion dark matter.

%

\vspace{0.1cm}
The quantum Hall effect (QHE) \cite{von, girvin} is observed in two-dimensional electron systems at low temperatures and under a strong magnetic field $B$ perpendicular to the two-dimensional surface. These systems are typically realized in semiconductor quantum wells, with the semiconductor sample known as a Hall bar.

In general, Hall conductivity $\sigma_{xy}$ varies with the magnetic field $B$. 
However, when $\sigma_{xy}$ is measured in a system at low temperatures ($\sim 1$K) and under a strong magnetic field 
($\sim 10$T), the Hall conductivity is quantized as $\sigma_{xy}=(e^2/2\pi ) \times n$, where $n$ is a positive integer. 
This quantization results in plateaus: regions of the magnetic field where $\sigma_{xy}$ remains constant despite changes in $B$.

The formation of these plateaus relates to the Fermi energy. While varying $B$ typically changes the electric current by altering the number of electrons contributing to it, the current remains constant on these quantized plateaus. When $B$ is further decreased  
(or increased) beyond a critical value, $\sigma_{xy}$ begins to increase (or decrease), leading to a transition to the next plateau, 
$\sigma_{xy}=(e^2/2\pi ) \times (n+1)$ ($\sigma_{xy}=(e^2/2\pi ) \times (n-1)$). 
This transition, called a plateau-to-plateau transition, occurs over a specific range of $\Delta B$, 
known as the transition width $\Delta B$.

\vspace{0.1cm}
In this paper, we aim to analyze these plateau transitions in detail, incorporating the potential effects of axions. Additionally, we use Hall resistance $\rho_{xy}$, which is equal to $1/\sigma_{xy}$ on the plateaus. On these plateaus, the diagonal resistances vanish 
($\rho_{xx}=\rho_{yy}=0$), while they do not vanish in the transition regions, 
where both $\rho_{xx}$ and $\rho_{yy}$ are nonzero.

\vspace{0.1cm}
The presence of the plateau can be understood by noting that, in general, two-dimensional electrons become localized due to a disorder potential $V$, which represents the effects of impurities, defects, and similar factors (hereafter referred to as disorder potential). However, under a strong magnetic field, some electrons remain delocalized.

In the absence of the disorder potential, electrons in a strong magnetic field form Landau levels with energies given by $E_n=\omega_c(n+1/2)$, where $\omega_c=eB/m_e^0$ represents the cyclotron energy 
and $m_e^0$ denotes the effective electron mass in the semiconductor. Each Landau level is highly degenerate, with the number of degenerate states per unit area equal to $eB/2\pi$. 

However, when a disorder potential $V$ is introduced, this degeneracy is lifted. Most electrons become localized, while some remain delocalized with energies close to $E_n$. Specifically, there exists an energy region around $E_n$	
in which states are not localized; we refer to this region as the mobility gap. Understanding the presence of the mobility gap is crucial for analyzing the plateau-to-plateau transitions.

\vspace{0.1cm}

In this paper, we discuss the mechanisms underlying plateau-to-plateau transitions and present compelling evidence for the presence of the axion by analyzing these transitions in previous experiments.
\vspace{0.1cm}

As the magnetic field $B$ is varied, Fermi energy varies. The Hall conductivity remains constant as long as the Fermi energy lies within localized states, which do not contribute to electric current. It is an insulator phase. 
This constancy in conductivity produces a plateau in $B$. 
A transition occurs only when the Fermi energy enters the extended states within the mobility gap, as these extended states can carry current. It is a metal phase. The transition ends once the Fermi energy exits the mobility gap.
Plateau-plateau transition can be seen as a metal-insulator transition. The transition is characterized by the transition width $\Delta B$,
within which metal-insulator transition arises.
\vspace{0.1cm}

The size of the mobility gap depends on factors such as temperature, the frequency of any external microwaves, and the dimensions of the Hall bar. The width $\Delta B$ is directly determined by the mobility gap. Therefore, to observe the axion effect in $\Delta B$, it is essential to examine how these factors influence the mobility gap.

\vspace{0.1cm} 

We examine in detail several experiments that demonstrate how the width behaves under various conditions: it decreases as the temperature (or the frequency of externally applied microwaves)
decreases, or as the size of the Hall bar increases.
Thus, $\Delta B$ depends on temperature, microwave frequency, and Hall bar size. We discuss these behaviors of $\Delta B$ by analyzing how the mobility gap varies with these factors.

\vspace{0.1cm}
Theoretically, the width $\Delta B$ is expected to vanish for an infinitely large Hall bar at zero temperature with no external radiation. However, in actual experiments using finite-sized samples, a non-vanishing width saturates at low temperatures or low frequencies, indicating a threshold below which $\Delta B$ does not decrease further even as temperature or frequency decreases. 
That is, $\Delta B$ decreases with the decrease of temperature or frequency of external microwave. But it saturates
at a critical temperature or frequency. It does not decrease even more as temperature or frequency decreases below a critical one.
Most of this saturation is due to finite-size effects, while a portion may result from the axion effect.

\vspace{0.1cm}
The key point in our discussion is that, even with an infinitely large Hall bar, the width $\Delta B$ remains non-zero at zero temperature due to the presence of axion-generated microwaves, albeit weak. Therefore, by carefully examining the behavior of
$\Delta B$ at low temperatures while varying the frequency of external microwaves and the size of the Hall bar, we can potentially detect a signature indicating the presence of the axion.

\vspace{0.1cm}
(It is well known \cite{sikivie, iwazaki01} that under a magnetic field $\vec{B}$,
axion dark matter generates an oscillating electric field $\vec{E}_a\propto \cos(m_at)\vec{B}$.
This field induces oscillating electric currents in nearby metals, producing axion microwaves. In quantum Hall effect experiments, metals exposed to the magnetic field $\vec{B}$ are always present around the Hall bar and are not necessarily oriented perpendicular to $\vec{B}$.
In this paper, we assume the presence of such axion-generated microwaves in quantum Hall effect experiments.)

\vspace{0.1cm}
As will be shown later, the effect of axions can be observed at low temperatures (\(< 30\) mK) and in sufficiently large two-dimensional electron systems with an area of \( S > 10^{-3} \, \rm{cm}^2 \). Otherwise, the effect is obscured by thermal influence. In other words, the lower the temperature and the larger the area \( S \), the more clearly the axion effect becomes visible.

\vspace{0.1cm}

In most of previous experiments \cite{engel, exp5, hohls, hohls2, balaban, exp6}, 
the observed saturation frequency $f_s$ is generally below $1$GHz at temperatures under $100$mK, largely because the Hall bar sizes were not sufficiently large to observe the axion effect or temperature is not sufficiently low. In order to observe the axion effect,
the larger size of Hall bar is needed for sufficiently large amount of the axion microwave to be absorbed and low temperature to suppress thermal noise.
However, in some of experiments \cite{exp8, exp9}, a higher saturation frequency of approximately \(2.4\) GHz than ever was observed at a low temperature of \(35\) mK using a Hall bar larger than any previously used. This result is unexpected, as saturation frequency typically decreases with lower temperatures and larger Hall bar sizes. We would thus anticipate a much lower saturation frequency, 
below $1$GHz, yet the observed frequency remains high at around $2.4$GHz.

\vspace{0.1cm}
We suggest that this anomalously high saturation frequency $f_s=2.4$GHz 
is due to the axion, with $f_s$ being theoretically given by $f_s=m_a/2\pi$. 
The axion effect is thus expected to manifest as a high saturation frequency $f_s\simeq 2.4$GHz at low temperature such as $35$mK when the Hall bar size is large.

\vspace{0.1cm}
Further evidence supporting the presence of the axion is provided by the experiments \cite{exp8, exp9}, which demonstrate that the saturation frequency $f_s\sim2.4$GHz is independent of the Hall bar size 
and that an almost identical frequency $f_s$ is observed in a different plateau transition. 
( The difference is characterized by different index, for instance, $n$ in Hall conductance $\sigma_{xy}=(e^2/2\pi ) \times (n+1)$. )
This transition differs from the one associated with the plateau transition width $\Delta B$ 
that saturates at $f_s=2.4$GHz, as noted above.

\vspace{0.1cm}

Additionally, an experiment \cite{exp4} using large Hall bar shows that the saturation temperature $T_s\sim 20$mK does not decrease as the size of the Hall bar increases. Generally, saturation temperature decreases with increasing Hall bar size due to finite-size effects. We discuss 
that this size-independent saturation temperature is caused by the axion effect, rather than a finite-size effect. Specifically, by calculating
contribution of the axion effect on the width $\Delta B$,
we show that the axion effect on $\Delta B$ dominates 
thermal effect at such low temperature. The thermal effect rapidly vanishes $\propto \exp(-m_a/T)$ as temperature $T$ decreases.
Thus,
the saturation temperature is almost independent of the Hall bar size or specific sample properties.

\vspace{0.1cm}
Interestingly, another experiment \cite{exp44} shows a similar axion effect, with the observed saturation temperature nearly identical to that in the previous experiment, despite using a different sample.

\vspace{0.1cm}

It is important to emphasize that the absorbed power of axion microwaves in the quantum Hall state is significantly enhanced. We have demonstrated \cite{iwa, iwa2} that this power scales with the square of the number of electrons in the state, where the electron number density is typically on the order of $10^{11}\rm cm^{-2}$. 
This is in contrast to resonant cavity experiments \cite{sikivie}, where the absorbed power is only proportional to the number of electrons on the cavity surface.

\vspace{0.1cm}
This enhancement effect in the quantum Hall state makes the axion signal potentially observable at temperatures 
below $30$mK when the Hall bar size exceeds $10^{-3}\rm cm^2$. 
The experiments that show signs of the axion effect meet these conditions.

\vspace{0.1cm}
To confirm the presence of the axion effect (i.e., axion-induced microwaves), we suggest shielding the Hall bar from axion microwaves and observing whether the high saturation frequency (approximately $2.4$GHz) disappears or decreases. 
In this paper, we propose an improved shielding method, which is more practical than the approach in our previous work.

\vspace{0.1cm}
We also outline additional confirmation methods. After detecting a saturation frequency of approximately $2.4$GHz at low temperatures, we can examine whether this frequency changes with temperature or Hall bar size. Furthermore, it would be informative to test whether the same saturation frequency $f_s\sim 2.4$GHz appears across different Hall bar samples.

\vspace{0.1cm}
In this paper, 
we use physical units, $c=1$, $k_B=1$ and $\hbar=1$. Our idea is also applicable to dark photon with frequencies of microwaves
discussed below.

\section{localized states and extended states}

We explain the integer quantum Hall effect as follows. Using semiconductors like GaAs or Si, a two-dimensional electron gas is formed, where electrons are confined in a quantum well that is broad in the horizontal direction (tens of micrometers or more) but narrow in the vertical direction (less than ten nanometers). As a result, electrons can move freely in the horizontal direction but are restricted in the vertical direction at low temperatures, where the energy spacing in this direction greatly exceeds both thermal energy (on the order of several tens of Kelvin) and any externally applied microwave energy.

When a perpendicular magnetic field $B$ is applied to the plane of the two-dimensional electron gas, electrons form quantized Landau levels. These levels are labeled by integer $n\ge 0$, with energies given by $E_n = \omega_c(n + 1/2)$, where the cyclotron energy $\omega_c = eB / m^0_e$, and $m^0_e$ is the effective electron mass (e.g., $m^0_e = 0.067 m_e$ in GaAs, where $m_e$ is the actual electron mass). The cyclotron energy $\omega_c$ is on the order of $10^{-2}(B / 10\text{T})$ eV, and each Landau level has a degeneracy of $eB / 2\pi$ states per unit area.

When including the Zeeman energy, each Landau level splits into two states with energies $E_{n\pm} = \omega_c (n + 1/2) \pm g \mu_B B$, where $g \approx 0.44$ and the Bohr magneton $\mu_B = e / 2m_e$. The Zeeman energy, of the order of $10^{-3}(B / 10\text{T})$ eV, is smaller than the cyclotron energy $\omega_c$. Thus, each spin-split Landau level retains the same degeneracy of $eB / 2\pi$.

In general, impurities and defects are present on the two-dimensional surface of the Hall bar, creating a disorder potential $V$ that affects the electrons. This potential is assumed to be much smaller than the cyclotron energy $\omega_c$, partially lifting the degeneracy of Landau levels. It is well known that most electrons become localized in these disordered regions and are unable to carry electric current. Only the states with energy $E_{n\pm}$ remain extended in an infinitely large Hall bar, allowing electrons in these states to conduct current.

\begin{figure}[htp]
\centering
\includegraphics[width=0.65\hsize]{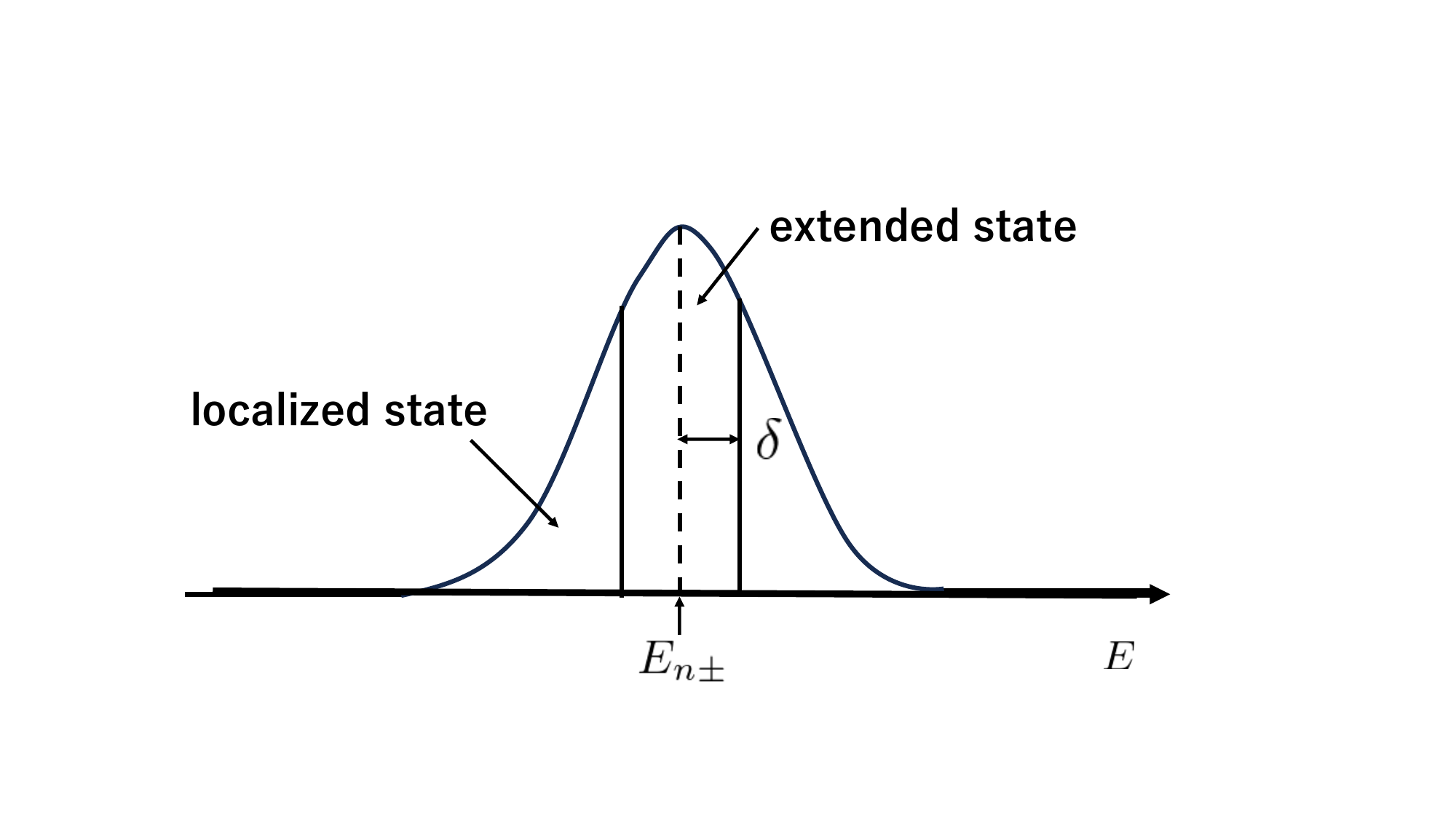}
\caption{density of state. $\delta$ represents mobility gap }
\label{1f}
\end{figure}

The coherence length $\xi(E)$, representing the spatial extent of a localized state with energy $E$, scales as $\xi(E) \sim |E - E_{n\pm}|^{-\nu}$ with $\nu \approx 2.4$ as $E \to E_{n\pm}$ \cite{aoki1,aoki2}. In a finite Hall bar, there exists an energy range $\delta \ge |E - E_{n\pm}|$ within which the size of the state exceeds the Hall bar dimensions, implying the presence of effectively extended states that can carry current. This energy range $\delta$, known as the mobility gap, is shown in Fig.(\ref{1f}). It vanishes in the limit of an infinitely large Hall bar, i.e., $\delta(L) \to 0$ as $L \to \infty$. When states within this gap are occupied, they contribute to electric current; otherwise, they do not. The mobility gap is crucial in plateau-to-plateau transitions.

Since the disorder potential $V$ is much weaker than the cyclotron and Zeeman energies, there is no mixing between Landau levels. Fig.1 depicts a typical density of states, assuming that $V$ contains roughly equal regions of positive and negative local potential energies.

It is worth noting that the width of the density of states is determined by the strength of the disorder potential $V$. For example, if we assume a density of states $\rho(E) \propto \sqrt{1 - ((E - E_{n\pm}) / \Delta E)^2}$ within the range $|E - E_{n\pm}| \le \Delta E$, the width $\Delta E$ is set by $V$, with a weaker $V$ leading to a narrower $\Delta E$. As shown in previous work \cite{iwa,iwa2}, axion-generated microwaves are absorbed more effectively in Hall bar samples with a weaker potential $V$. Thus, the axion effect may be more pronounced in samples with a smaller $\Delta E$.

\vspace{0.1cm}
The range of the disorder potential $V$ plays a crucial role in determining its strength. Generally, samples with a significant short-range potential $V_s$ exhibit a large $\Delta E$, while those dominated by a long-range potential $V_l$ tend to have a smaller $\Delta E$. Consequently, the axion effect is more likely to be observable in samples primarily influenced by long-range disorder potential.

Short-range disorder potential causes large-angle scattering of electrons, resulting in a shorter electron relaxation time, which in turn leads to a larger $\Delta E$.

\vspace{0.1cm}

As long as the Fermi energy $E_f$ remains below $E_{n\pm} - \delta$, the Hall conductivity takes the quantized value $\sigma_{xy} = \frac{e^2}{2\pi} \times n$. As $E_f$ exceeds $E_{n\pm} - \delta$, $\sigma_{xy}$ gradually increases toward $\frac{e^2}{2\pi} \times (n + 1)$. Eventually, once $E_f$ surpasses $E_{n\pm} + \delta$, the Hall conductivity reaches $\sigma_{xy} = \frac{e^2}{2\pi} \times (n + 1)$ and stabilizes at this value, marking the beginning of the next plateau.

This change in conductivity is the plateau-to-plateau transition. Electrons in localized states with energies $E < E_{n\pm} - \delta$ do not contribute to electric current, so the Hall conductivity remains unchanged. It is insulator phase.
However, as extended states with energies $E$ within the range $\delta < |E_{n\pm} - E|$ become occupied, current flows, leading to an increase in $\sigma_{xy}$. It is metal phase.
Once $E_f$ exceeds $E_{n\pm} + \delta$, electrons begin occupying localized states again, causing the Hall conductivity to stabilize.

Additionally, it is important to note that the Fermi energy $E_f$ shifts with the magnetic field $B$. As $B$ increases, the degeneracy $eB/2\pi$ of each Landau level also increases, causing a decrease in $E_f$, since the electron density remains constant while the degeneracy changes. When $E_f$ transitions from $E_{n\pm} - \delta$ to $E_{n\pm} + \delta$, the plateau-to-plateau transition completes.

We define the width $2\Delta B$ of the magnetic field as the range corresponding to the Fermi energy width $\Delta E_f = 2\delta$, spanning from $E_{n\pm} - \delta$ to $E_{n\pm} + \delta$.

%
%
%

\section{electron distribution}

%

At zero temperature, the electron distribution has a sharp boundary: all states with energies below the Fermi energy $E_f$ are occupied, while all higher-energy states are empty. At finite temperatures ($T \neq 0$), however, this distribution becomes smeared around the chemical potential $\mu_c$. (At low temperatures, typically below 1 K, the chemical potential closely approximates the Fermi energy, so we refer to it as the Fermi energy even for $T \neq 0$.)

\begin{figure}[htp]
\centering
\includegraphics[width=0.65\hsize]{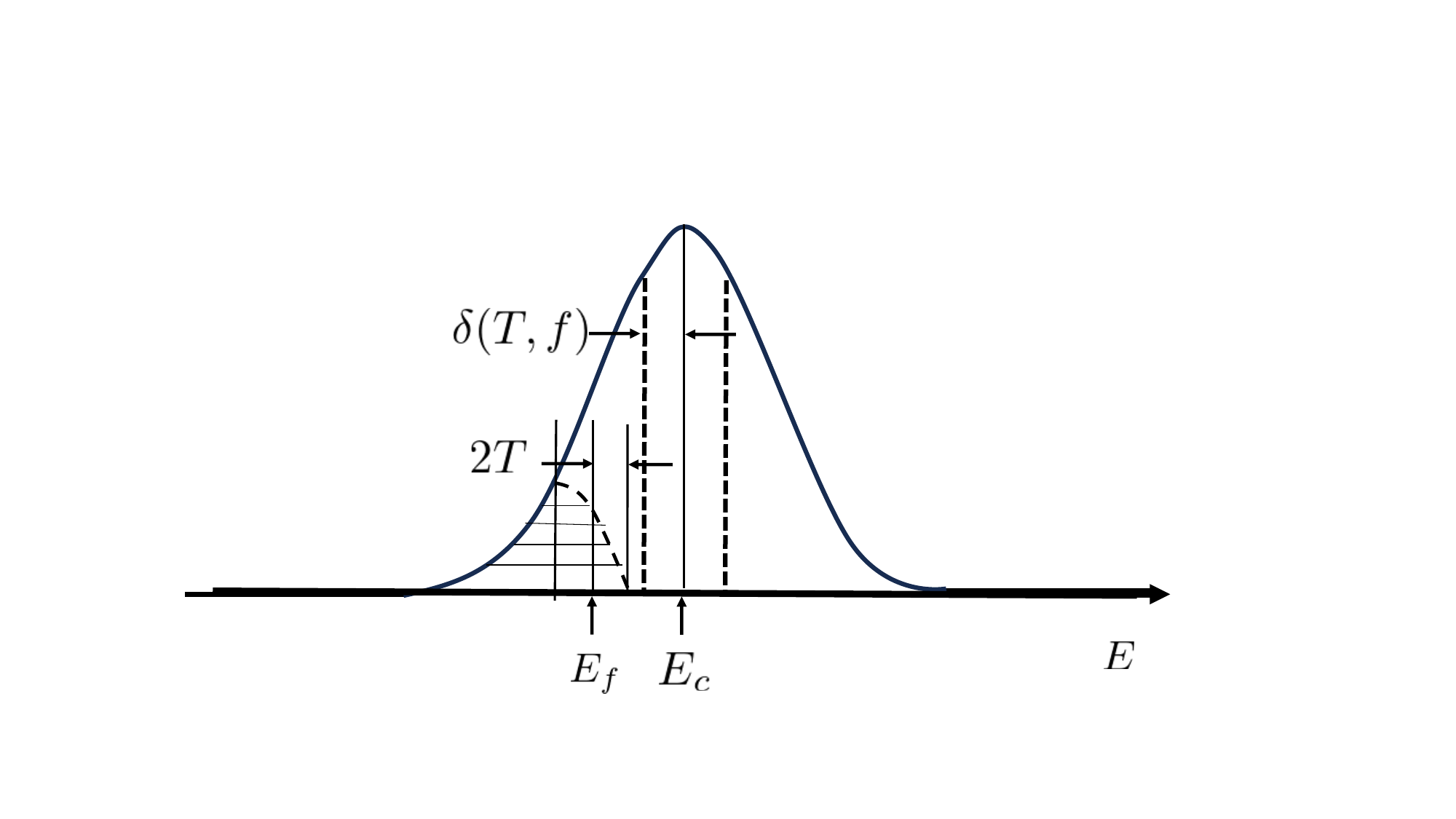}
\caption{The electron distribution is broadened by temperature, while a microwave broadens it with \( 2T \) replaced by \( 4\pi f \). Additionally, we illustrate the mobility gap \( \delta(T, f) \). }
\label{2f}
\end{figure}

While the electron distribution does not have a sharp boundary at finite temperatures, it can be approximated as having a sharp boundary, where states with energies above $E_f + 2T$ are effectively empty. In other words, the distribution is spread over a range of approximately $4T$ around the Fermi energy $E_f$ (see Fig.(\ref{2f})). With this simplified view, the plateau-to-plateau transition at $T \neq 0$ becomes easier to understand.

When the Fermi energy is below $E_{n\pm} - \delta - 2T$, the Hall conductivity remains on a plateau. As $E_f$ rises and crosses $E_{n\pm} - \delta - 2T$, the transition begins, causing an increase in Hall conductivity. When $E_f$ reaches $E_{n\pm} + \delta + 2T$, the conductivity stabilizes at the next plateau, completing the transition. Therefore, at finite temperatures, the Fermi energy range $\Delta E_f$ associated with this transition is given by $\Delta E_f = 2\delta + 4T$. 
Experimentally, we observe the transition width $\Delta B$, which is determined by the width $\Delta E_f$. When we examine the width
$\Delta B$, we may instead discuss $\Delta E_f$.

\vspace{0.1cm}
We have observed that the transition width $\Delta B$ also depends on the frequency of external microwaves, showing a similar pattern to its temperature dependence: $\Delta B \propto f^{\gamma}$ \cite{exp10} and $\Delta B \propto T^{\kappa}$ \cite{deltaB}, with $\gamma, , \kappa \sim 0.4$ as long as $T$ or $f$ remains above a critical value. Below this critical threshold, the width $\Delta B$ saturates. In this context, frequency $f$ is analogous to temperature $T$.

This similarity can be explained as follows. Electrons with energy $E$ that absorb microwave radiation of frequency $f$ transition to states at energy $E + 2\pi f$, and subsequently lose energy through collisions with phonons. Thus, the external microwave causes the electron distribution to spread in a manner similar to that induced by temperature, where $T \propto 2\pi f$. Although the smearing due to microwaves differs from that due to thermal effects, it is reasonable to approximate the electron distribution as also having a sharp boundary at $E = E_f + 4\pi f$ at $T = 0$, where states with energy greater than $E_f + 4\pi f$ are empty. The simplification makes
plateau-plateau transition easy to understand.

This form of the distribution does not necessarily match the thermal distribution at $T = 2\pi f$, but we assume the presence of a sharp cutoff at $E = E_f + 4\pi f$. Thus, at $T = 0$ but in the presence of an external microwave of frequency $f$, the plateau transition width $\Delta E_f$ can be approximated as $\Delta E_f = 2\delta + 8\pi f$. It is important to note that the frequency $f$ defining the distribution cutoff is not necessarily the actual microwave frequency, although it is close to this value.

\begin{figure}[htp]
\centering
\includegraphics[width=0.65\hsize]{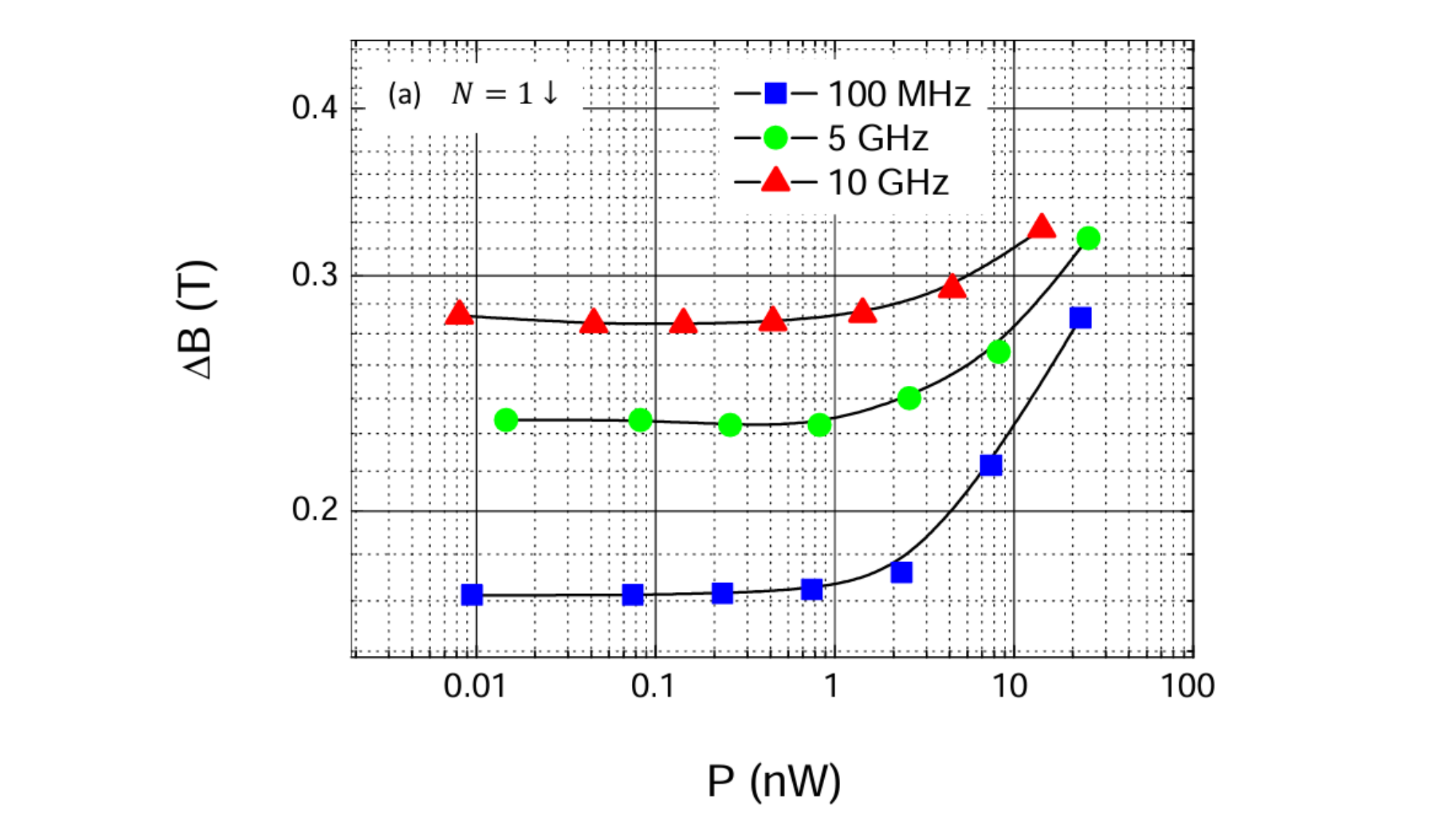}
\caption{dependence of $\Delta B$ on the power of microwave. Saturation value only depends on the frequency of microwave, not its power. Ref.\cite{exp9}}
\label{2fa}
\end{figure}

We note that the microwave power used in experiments must be kept sufficiently low to avoid changing the sample temperature. In practice, microwaves with such low power are employed to ensure that the width $\Delta B$ or $\Delta E_f$ can saturate. Generally, this width depends on the microwave power, decreasing with reduced power until it reaches saturation below a specific threshold value \cite{exp9};
see Fig.(\ref{2fa}) .  The saturation width $\Delta B$ only depends on the frequency, not the power of microwave.

This is important for the determination of $\Delta B$ under two microwaves with different frequencies.
The microwave generated by axion has much lower power than external one imposed, but the axion microwave dominates
the external one for the determination of $\Delta B$ when the frequency $m_a/2\pi $ is higher than the 
frequency $f$ of external microwave. 
This is because the width $\Delta B$ is determined by electrons in extended states near the mobility edge and such electrons
are excited from localized states by microwave with higher frequency. Even if the power is stronger than the one
of the axion microwave, electrons within the mobility gap are not produced by the external microwave as far as their frequency $f$
is not sufficiently large for electrons to be excited to the mobility gap. The phenomena is similar to photoelctric effect.
In the later section we compare the number of electrons in the mobility gap exited by axion microwave and thermal effect.
We find that the axion effect is dominant for the thermal effect at low temperature of the order of $10$mK.
Thus, the axion microwave dominates over both the external microwave and the thermal effect at such a low temperature.

%
%
%

\vspace{0.1cm}
The electron distribution behaves in the following when we change microwave frequency but fix temperature.
When $2\pi f \gg T$, microwave clearly dominates so that the cut off in the distribution is given by $4\pi f$. 
It follows that  $\Delta E_f=2\delta+8\pi f$.
On the other hand,
when $T \gg 2\pi f$, temperature dominates so that the cut off is given by $2T$. It follows that  $\Delta E_f=2\delta+4T$.
There is a critical frequency $f_c=T/2\pi$ 
below which the temperature dominates to determine the cut off.

\vspace{0.1cm}

In the discussion we mention that temperature is dominant for $f < f_c = T/2\pi$, while microwaves dominate when $f > f_c$. Here, $f_c = T/2\pi$ serves as an approximate threshold frequency rather than an exact frequency. 
The frequency and the temperature used as cutoffs in the electron distribution are approximate ones
so that the relation $f_c=T/2\pi$ does not represent exact relation between temperature and frequency. 
The smooth out of electron distribution by thermal effect is essentially different from 
the one by external microwave effect.

\vspace{0.1cm}
In the discussion below, we explore which of two microwaves 
with different frequencies, $f_1$ and $f_2$, takes precedence when both are present. 
( One is axion microwave and the other is external microwave imposed. )
In this case, when $f_2 > f_1$, the microwave with frequency $f_2$ is dominant, while for $f_2 < f_1$, $f_1$ is dominant. The transition between dominance by $f_1$ and $f_2$ occurs precisely at $f_1 = f_2$, as both microwaves affect the electron distribution in a comparable manner even if their powers are different. Only frequency is relevant for precedence in determination of
the width $\Delta B$.
This differs from the previously mentioned approximate relation $f_c = T/2\pi$.
In latter section we use the relation $f_1 = f_2$ to obtain the axion mass $m_a$, because the axion generates microwave with
frequency $m_a/2\pi$.

\section{mobility gap}

The mobility gap represents the energy range $\delta$ over which states are effectively extended, with energy $E$ satisfying $\delta \ge |E - E_c|$, where $E_c = E_{n\pm}$. States within this gap contribute to electric conduction, causing the Hall conductivity to increase as the occupation number in these states grows. The mobility gap also depends on the spatial extent of the two-dimensional electron system, specifically the Hall bar size $L$. As discussed earlier, the coherence length $\xi(E)$, or the spatial extent of a state with energy $E$, behaves as $\xi(E) \propto |E - E_c|^{-\nu}$ with $\nu \simeq 2.4$ as $E \to E_c$. At zero temperature ($T = 0$), the mobility gap $\delta(L)$ is defined such that states with energies $E$ satisfying $\delta \ge |E - E_c|$ extend beyond the size $L$, enabling them to carry electric current. Importantly, this gap $\delta(L)$ vanishes as $L \to \infty$.

The mobility gap is influenced by the disorder potential. When the disorder potential $V$ is stronger, a greater proportion of electrons become localized, resulting in a shorter localization length. Consequently, the mobility gap generally decreases as the disorder potential strengthens. Simultaneously, the density of states $\rho(E)$ broadens with a stronger disorder potential. For example, if $\rho(E) \propto \sqrt{1 - ((E - E_{n\pm}) / \Delta E)^2}$, the width $\Delta E$ increases with disorder strength. A useful measure of disorder strength is the mobility $\mu$, commonly used in semiconductor physics, and defined by $\sigma = e \rho \mu$, where $\sigma$ is the electrical conductivity and $\rho$ is the electron density. High mobility implies a weaker disorder potential, which corresponds to a larger $\delta$ and a smaller $\Delta E$, while low mobility indicates a stronger potential, yielding a smaller $\delta$ and a larger $\Delta E$. (Note that $\Delta E > \delta$ always holds.)

\vspace{0.1cm}
Now let us consider the thermal effect on the mobility gap. The phase coherence length represents the spatial region over which quantum coherence is maintained. At $T = 0$, this length is infinite, but it decreases with increasing temperature due to thermal fluctuations. As long as the phase coherence length exceeds the size $L$ of the Hall bar, electric current flows in a manner similar to that in metals, with electrons in extended states contributing to current according to the standard conduction mechanism. However, when the phase coherence length becomes shorter than $L$, the electric current is not carried by the standard conduction mechanism. 
The current is instead carried through thermally assisted hopping conduction. In this mechanism, a localized electron with energy $E$ can thermally hop to a neighboring localized state with energy $E'$ within an energy range $|E - E'| \le T$. As a result, the effective mobility gap increases with temperature, especially beyond a critical temperature $T_c$, where the phase coherence length equals the Hall bar size $L$. For $T \le T_c$, we have $\delta(T) = \delta(T=0)$.

Thus, the mobility gap depends on both temperature and the Hall bar size $L$. Since $\delta(T=0)$ vanishes as $L \to \infty$, the critical temperature $T_c$ also vanishes in this limit.

That is, we have

\begin{equation}
\delta (T')> \delta (T) \quad \mbox{for} \quad T'>T \ge T_c,\,\,\,\,
\delta (T)=\delta (T=0) \quad \mbox{for} \quad T\le T_c, \quad \mbox{and} \quad T_c \to 0 \quad \mbox{as} \quad  L\to \infty
\end{equation}

\vspace{0.1cm}
Similarly, the mobility gap $\delta(f)$ depends on the microwave frequency $f$. Beyond a critical frequency $f_c$, the gap $\delta(f)$ increases with frequency for $f \ge f_c$. Conversely, for frequencies $f \le f_c$, we have $\delta(f) = \delta(f = 0)$. This critical frequency $f_c$ is dependent on the system size $L$ and approaches zero as $L \to \infty$.
Thus, we have

\begin{equation}
\delta (f')> \delta (f) \quad \mbox{for} \quad f'>f \ge f_c,\,\,\,\,
\delta (f)=\delta (f=0) \quad \mbox{for} \quad f\le f_c, \quad \mbox{and} \quad f_c \to 0 \quad \mbox{as} \quad  L\to \infty
\end{equation}

\vspace{0.1cm}
In general, mobility gap depends on both temperature and frequency of microwave. It is reasonable to suppose that

\begin{eqnarray}
\delta (T,f)&=&\delta (T=0,f) \quad \mbox{for} \quad T\le T_c(f), \quad \mbox{and} \quad T_c(f') \ge T_c(f) \quad \mbox{for} \quad f'\ge f \\
\delta (T,f)&=&\delta (T,f=0) \quad \mbox{for} \quad f\le f_c(T), \quad \mbox{and} \quad f_c(T')\ge f_c(T) \quad \mbox{for} \quad T'\ge T
\end{eqnarray}

\section{plateau-plateau transition}

Using these results, we discuss how plateau-plateau transitions occur under finite temperature and external microwaves. We examine the transition width $\Delta B$, determined by the width $\Delta E_f$ of the Fermi energy over which the transition happens.

The key question is how the width $\Delta E_f$ is influenced by temperature and microwave effects. This width is defined as the energy difference $E_2 - E_1$ between points $E_1$ and $E_2$, where the transition begins at $E_f = E_1$ and ends at $E_f = E_2$, passing through the center energy $E_c = E_{n\pm}$ as the magnetic field decreases. From Fig.(\ref{2f}), we find $E_1 = E_c - 2T - \delta(T, f)$ and $E_2 = E_c + 2T + \delta(T, f)$, leading to $\Delta E_f = 2\delta(T, f) + 4T$. This corresponds to the case where temperature primarily determines the electron distribution cutoff. Conversely, in the case where the microwave dominates, we have $\Delta E_f = 2\delta(T, f) + 8\pi f$, where $f$ is the external microwave frequency.

\begin{figure}[htp]
\centering
\includegraphics[width=0.65\hsize]{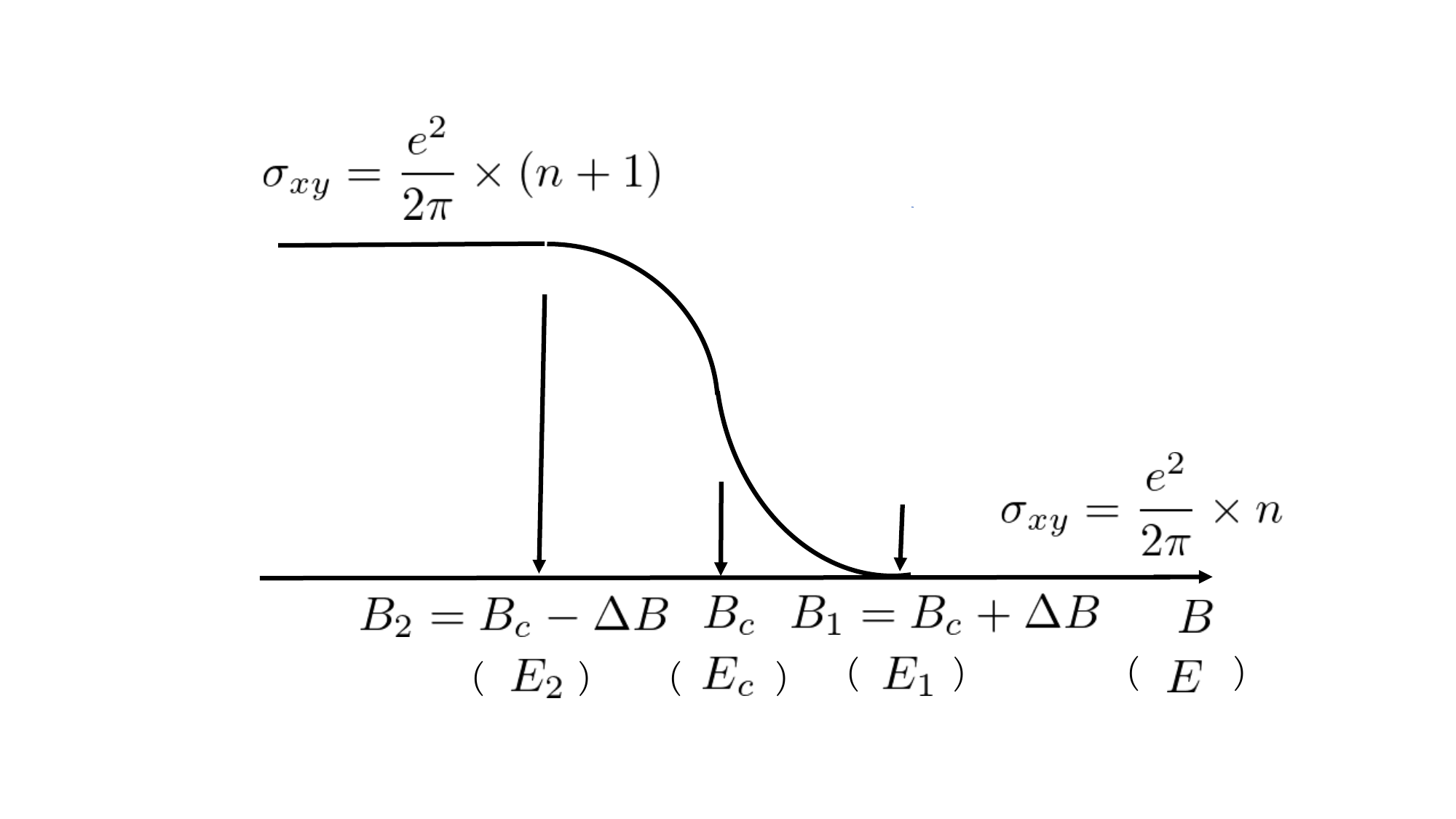}
\caption{Plateau-plateau transition. We define $\Delta B$ as illustrated in the figure. }
\label{3f}
\end{figure}

\vspace{0.1cm}
As the magnetic field $B$ decreases, the Fermi energy increases. The transition initiates at $B = B_1$ and concludes at $B = B_2$, with the transition width $\Delta B$ defined as $2\Delta B = B_1 - B_2$ (see Fig.~(\ref{3f})). The observed $\Delta B$ is thus set by $\Delta E_f = 2\delta(T, f) + 4T$ or $\Delta E_f = 2\delta(T, f) + 8\pi f$. Here, the first term $\delta(T, f)$ represents the mobility gap, while the second term reflects the energy cutoff in the electron distribution: $2T$ for temperature dominance, and $4\pi f$ for microwave dominance. When $T \ll 2\pi f$, the microwave cutoff energy $4\pi f$ is dominant; when $T \gg 2\pi f$, the temperature cutoff energy $2T$ prevails. In the case where $T \sim 2\pi f$, the dominant factor shifts between temperature and microwave effects. It should be noted that the condition $T = 2\pi f$ serves as an approximate threshold for the transition, which can occur, for example, at frequencies close to $T / 2\pi$.

\vspace{0.1cm}
In some cases, the transition is characterized by the derivative $d\rho_{xy}/dB$. As the transition progresses from $\rho_{xy} = 2\pi / e^2 \times 1/n$ to $\rho_{xy} = 2\pi / e^2 \times 1/(n+1)$, we can approximate $d\rho_{xy}/dB \approx \Delta \rho_{xy} / \Delta B$ with $\Delta \rho_{xy} = 2\pi / e^2 \times 1 / (n(n+1))$. Therefore, the behavior of $d\rho_{xy}/dB$ as a function of temperature $T$ or frequency $f$ reflects the behavior of $\Delta B$.

\vspace{0.1cm}
It is worth mentioning that even without an externally applied microwave, a microwave component due to axion dark matter may be present in experiments. This microwave frequency $f_a$ is given by the axion mass, $f_a = m_a / 2\pi$. 
The axion-generated microwave is very weak, but as shown in previous studies (and also in the following section), the axion effect can become evident at low temperatures, $\le 50$ mK, and for large Hall bars with surface area $\ge 10^{-3} , \mathrm{cm}^2$. Low temperatures help the axion effect dominate over thermal background radiation, while a large surface area enables sufficient absorption of radiation energy from the axion by two-dimensional electrons for the axion effect to be observable.

Thus, at higher temperatures ($> 50$ mK) or for smaller Hall bars, the axion-generated microwave plays no significant role in the plateau-plateau transition. The thermal effect makes the axion effect not to be observed.
At sufficiently low temperatures, e.g., $T \le 50$ mK, applying an external microwave with frequency $f$ 
can reveal a critical frequency $f_s$, below which the transition width $\Delta B(T, f)$ saturates with respect to $f$. This critical frequency $f_s$ differentiates between dominance by the external microwave and the axion-induced microwave. When $f > f_s$, the external microwave dominates the electron distribution, while the axion microwave prevails when $f < f_s$. 
The critical frequency $f_s$ is given directly by the axion mass: $f_s = m_a / 2\pi$. 
As we have mentioned in previous section, even if the power of the axion microwave is much lower 
than the one of the external microwave, the microwave with higher frequency is dominant for the determination of the width $\Delta B$.
Electrons within the mobility gap are not excited from localized states when external microwaves have no sufficient high frequency to
produce such electrons. The presence of such electrons determines the width.

\section{axion contribution}
\label{axion}


Here, we compare the contributions of the axion effect and thermal effects during the plateau-to-plateau transition. As the Fermi energy \(E_f\) increases and approaches the mobility edge \(E_c - \delta\), some electrons occupy extended states, which can result from either thermal effects or the absorption of microwaves produced by the axion. In other words, electrons in the tail of the thermal distribution occupy extended states, while electrons that transition from localized states due to axion-induced microwave absorption also occupy extended states. Clearly, at high temperatures, thermal electrons dominate over those induced by the axion, making the thermal effect larger than the axion effect. However, as the temperature decreases, the thermal effect becomes smaller and eventually, the axion effect becomes dominant. Our goal is to determine the critical temperature below which the axion effect takes precedence.
\vspace{0.1cm}

When determining $\Delta B$, an important factor is that the Fermi energy changes with variations in the magnetic field. The key point is at which value of the magnetic field electrons transit  within the mobility gap. In other words, it corresponds to the magnetic field value at which the plateau transition begins. The difference between this starting value and the value at which the transition ends is $2\Delta B$.  

Drawing an analogy to semiconductors, this is similar to determining how much the Fermi energy, located in the valence band (localized states), needs to increase for electrons in the valence band to be excited to the conduction band (delocalized states) due to the influence of microwaves. The Fermi energy level that allows electrons to be excited to the conduction band by absorbing microwaves determines $\Delta B$.  

In this analogy, the quantum Hall state corresponds to a semiconductor with no energy gap between the valence band (localized states) and the conduction band (delocalized states).

\vspace{0.1cm}
At this point, it is crucial to consider that thermal effects can also excite electrons into the conduction band. However, thermal effects are highly sensitive to temperature. The competition between microwave frequency and temperature determines which primarily excites the conduction electrons. Clearly, at sufficiently low temperatures, thermal effects are negligible. As the temperature increases, there is a certain threshold at which thermal effects surpass the influence of microwaves.

\vspace{0.1cm}
In the evaluation of the axion contribution, we assume 
an axion dark matter density $\rho_d$ of approximately $0.3\rm GeV/cm^3$ and use the QCD axion parameter $g_{a\gamma\gamma}$, defined as $g_{a\gamma\gamma} = g_{\gamma} \alpha / (\pi f_a)$, where the fine-structure constant $\alpha \simeq 1/137$ and the axion decay constant $f_a$ satisfies $f_a m_a \simeq 6 \times 10^{-3} \rm GeV^2$. The constant $g_{\gamma}$ takes values $g_{\gamma} = 0.37$ for the DFSZ axion model and $g_{\gamma} = -0.96$ for the KSVZ axion model. These standard parameters e.g. $g_{a\gamma\gamma}$
arise in QCD axion interaction with electromagnetic field \cite{iwa,iwa2}.

\vspace{0.1cm}
We suppose that Fermi energy is given such that $E_f=E_c-\delta-m_a+\delta E$ with $m_a \gg \delta  E>0$.
Thus, when $T=0$,
electrons excited with the absorption of axion microwave occupy the delocalized states with energies $E_{\beta}$ in a small range $E_c-\delta \le E_{\beta}\le E_c-\delta +\delta E$ in the mobility gap. That is, the plateau transition takes place at
Fermi energy $E_f=E_c-\delta-m_a$ at $T=0$. 
However, at sufficiently low temperatures, 
the energy of the excited electrons does not significantly exceed \( E_c - \delta + \delta E \).
It implies that when $E_f=E_c-\delta-m_a$, plateau transition begins even at non vanishing 
temperature $T \ll m_a$. 

\vspace{0.1cm}
At such temperatures, 
we evaluate number of electrons transiting from localized state to delocalized state by absorbing axion microwave described
by gauge potential $\vec{A}_a$. The energy of the microwave is given by the axion mass $m_a$.
Using the transition amplitude of electron in the localized state $\alpha$ to the delocalized state $\beta$

\begin{equation}
<\beta |H_a|\alpha>=
i(E_{\beta}-E_{\alpha})e\vec{A}_a\cdot <\beta|\vec{x}|\alpha>\equiv i(E_{\beta}-E_{\alpha})e\vec{A}_a\cdot \vec{L}_{\alpha\beta}
\end{equation}
with $H_a=\frac{-ie\vec{A}_a\cdot\vec{P}}{m_e^{\ast}}$ and $\vec{L}_{\alpha\beta}$( $\equiv <\beta|\vec{x}|\alpha>$,
we obtain number of electrons per unit time $\delta \dot{N}_a$ transiting to delocalized states
with energy $E_{\beta}$,

\begin{eqnarray}
\label{8}
\delta \dot{N}_a&=&2\pi S^2\int_{E_c-\delta -m_a}^{E_c-\delta } 
dE_{\alpha}\rho({E_{\alpha}+m_a})\rho(E_{\alpha})\big(1-f(E_{\alpha}+m_a)\big)f(E_{\alpha})m_a^2\Big(e\vec{A}_a\cdot \vec{L}_{\alpha\beta}\Big)^2 \\
&\simeq& 2\pi S^2m_a^2\Big(\frac{eB}{2\pi}\frac{2}{\pi \Delta E}\Big)^2e^2A_0^2A^2l_B^2 \times T\log 2 \\
&\sim & 0.83\times 10^{5}s^{-1}
\Big(\frac{A}{10^4}\Big)^2 \Big(\frac{S}{10^{-3}\mbox{cm$^2$}}\Big)^2
\Big(\frac{0.5\times10^{-4}\mbox{eV}}{\Delta E}\Big)^2
\Big(\frac{\rho_d}{0.3\,\mbox{GeVcm$^{-3}$}}\Big)\Big(\frac{B}{10\rm T}\Big)^3\Big(\frac{m_a}{10^{-5}\mbox{eV}}\Big)\Big(\frac{g_{\gamma}}{1.0}\Big)^2 \Big(\frac{T}{m_a}\Big) ,
\end{eqnarray}
where $f(E)=(1+\exp\big((E-E_f)/T\big) )^{-1}$ denotes Fermi distribution and  
$S$ does surface area of two dimensional electrons. ( We have taken $E_f=E_c-\delta-m_a+\delta E$ with $T\gg \delta E$.
When  $T\ll \delta E$, the last factor $T/m_a$ in the formula of $\delta \dot{N}_a$ is replaced by $\delta E/m_a$. )
We have put  $(e\vec{A}_a\cdot \vec{L}_{\alpha\beta})^2\equiv e^2A_0^2A^2l_B^2$ with $A_0=g_{a\gamma\gamma}a_0B/m_a$
( axion dark matter density $\rho_d=a_0^2m_a^2/2$ ).
We have replaced  $L_{\alpha\beta}$ by the average one $\overline{L_{\alpha\beta}}=Al_B$ 
( $l_B\simeq 8.2\times 10^{-7}\rm cm \sqrt{10T/B}$ ).
We explicitly use the formula of the density of state $\rho(E)$,

\begin{equation}
\rho(E)=\rho_0\sqrt{1-\Big(\frac{E-E_c}{\Delta E}\Big)^2} \quad \mbox{with} \quad |E-E_c|\le \Delta E 
\quad \mbox{otherwise} \quad \rho(E)=0
\end{equation} 
with $\rho_0=(eB/2\pi)\times 2/(\pi \Delta E)$,
where $\int dE \rho(E)=eB/2\pi$ represents the number density of electrons in a Landau level;
$\int dE \rho(E)\simeq 2.4\times 10^{11}\rm cm^{-2}$$(B/10T)$. We have assumed $(\delta+m_a)/\Delta E \ll 1$
so that the integral over $E_{\alpha}$ is trivial because approximately $\rho(E)\sim \rho_0$.

\vspace{0.1cm}
The width $\Delta E$ has been assumed to be $0.5\times 10^{-4}$eV. 
This value is found to be reasonable 
when using the approximation formula \( \Delta E = \sqrt{2\omega_c/(\pi \tau)} \), where \( \tau \) represents the relaxation time. The relaxation time may be estimated using the semiconductor mobility \( \mu \) 
as \( \tau = \mu m_e^0 / e \). By using \( \mu = 5 \times 10^5 \, \text{cm}^2/\text{Vs} \), we obtain a value that closely matches the one adopted here. The actual value may vary depending on the sample, but the difference is at most an order of magnitude.

\vspace{0.1cm}
We have supposed $m_a=10^{-5}$eV. The value is strongly suggested from an experiment mentioned later.
We have also supposed $A=10^4$. It is overlapping region $\overline{L_{\alpha\beta}}=Al_B$ 
between a localized state 
with energy $E_{\alpha}=E_c-\delta -m_a$ and delocalized state with energy $E_{\beta}=E_c-\delta $.
We note that the evaluation is performed under the limit $\delta E \to 0$.
Using the scaling law of coherence length mentioned previuosly, we estimate  
the ratio of the localization scale between both states such that  
$\xi(E_{\alpha})/\xi(E_{\beta})=|\delta +m_a|^{-2.4}/|\delta|^{-2.4}=
|1+m_a/\delta|^{-2.4}\simeq 0.07$ with $\delta=0.5\times 10^{-5}$eV or $\xi(E_{\alpha})/\xi(E_{\beta})\simeq 0.11$
with $\delta= 0.7\times 10^{-5}$eV. ( Note that $\Delta E>\delta $. ) 
When a Hall bar is a rectangle with a side ratio of $4:1$, 
$S=10^{-3}\rm cm^2\simeq (6.3\times 10^{-2}\rm cm) \times (1.8\times 10^{-2} cm)$.
Thus, $\xi(E_{\alpha})\simeq 6.3\times 10^{-2}$cm so that $\xi(E_{\beta})\sim 6\times 10^{-3}$cm.
The overlapping region we supposed is $Al_B=8.2\times 10^{-3}$cm$\sqrt{10T/B}$.
It is comparable to the localization length of the state $\alpha$.
It is the reason why we take the value $A=10^4$.
\vspace{0.1cm}

The value obtained here represents the number of electrons excited per unit time into the mobility gap by axion-induced microwaves. These electrons subsequently emit electromagnetic waves or phonons and return to localized states within an extremely short time.  

Additionally, when a current is applied in the \(x\)-direction during measurements, a voltage arises in the \(y\)-direction. As a result, electrons within the mobility gap flow in the \(y\)-direction and occupy edge states. In this scenario, how many electrons reach the edge states along the \(y\)-axis before returning to a localized state after excitation? By comparing the lifetime of electrons within which they return to localized states with the time required for them to flow along the \(y\)-axis, we can determine how many electrons are effectively excited into the mobility gap and contribute to voltage generation in the \(y\)-direction.  
( Without such electrons excited to mobility gap, the electric conductivity $\sigma_{xy}$ does not change so that it
stays in a plateau. ) 

\vspace{0.1cm}
Here, we first calculate the lifetime of electrons before they return to localized states via electromagnetic wave emission. (The effect of phonon emission is assumed to be comparable to that of electromagnetic waves.)  

Electrons excited near the mobility edge can transition to localized states up to the Fermi energy \(E_f\), as states below \(E_f\) are already occupied when $T\ll m_a$.
The transition probability per unit time for the emission of electromagnetic waves with energy \(E\) is given by 
$4\alpha A^2l_B^2E^3/3$ \cite{iwa}. Thus, the transition probability per unit time ( $\equiv \tau^{-1}$ ) for 
the emission of electromagnetic waves with energy \(E\), $0\le E\le \Delta E-\delta$ ( $\Delta E-\delta >m_a $ ) is

\begin{eqnarray}
\tau^{-1}&=&\frac{4\alpha}{3}A^2l_B^2\int_0^{\Delta E-\delta} E^3 S\rho(E_c-\delta-E)\big(1-f(E_c-\delta-E)\big) dE \\
&\simeq& \frac{\alpha}{3}A^2l_B^2 m_a^4 S\rho_0\simeq 1.8 \times 10^{10}\rm s^{-1} \Big(\frac{A}{10^4}\Big)^2
\Big(\frac{S}{10^{-3}\rm cm^2}\Big)\Big(\frac{0.5\times 10^{-4}\rm eV}{\Delta E}\Big).
\end{eqnarray}
under the assumption $\Delta E \gg m_a$.
It leads to the life time $\tau\simeq 5.6\times 10^{-11}s$ of electrons staying mobility gap.

\vspace{0.1cm}
Next, we calculate the time it takes for an electron excited within the mobility gap to flow out in the \(y\)-direction.
The electron occupies an edge state.  

When a current of $I_x=10^{-9}A$ is applied in the \(x\)-direction, a voltage \(V_y=\rho_{yx}I_x\) is generated in the \(y\)-direction. This voltage accelerates the electron, allowing us to determine the time $t_p$ it takes to traverse the sample's width \(w\) in the \(y\)-direction. We may roughly estimate average velocity of electrons using mobility $\mu$ of the sample
such that $v=\mu V_y/ew=\mu \rho_{yx}I_x/ew$.
We obtain 

 \begin{equation}
t_p=w/v=\frac{e w^2}{\mu \rho_{xy}I_x}\simeq 4.1\times 10^{-8}\mbox{s}
\Big(\frac{w}{10\mu\rm m}\Big)^2\Big(\frac{3\times 10^5\rm cm^2/Vs}{\mu}\Big)\Big(\frac{10^{-9}A}{I_x}\Big),
\end{equation}
with $\rho_{xy}=2\pi/e^2$. We note that the values of electric current $10^{-9}A$, mobility $\mu=3\times 10^5\rm cm^2/Vs$ and width $w=10\mu \rm m$ are 
ones typically used in experiments.

From these results, the fraction of electrons excited from localized states into the mobility gap by axion-induced microwaves that contribute to the actual Hall conductivity measurement is given by \(\tau / t_p\simeq 1.3\times 10^{-3}\).
Namely, the effective number of electrons per unit time excited from localized states to contribute the width $\Delta B$ is

\begin{equation}
\delta \dot{N}_a^{\rm eff}=\delta \dot{N}_a\Big(\frac{\tau}{t_p}\Big)\simeq 1.1\times 
10^{2}s^{-1}
\Big(\frac{S}{10^{-3}\mbox{cm$^2$}}\Big)
\Big(\frac{0.5\times10^{-4}\mbox{eV}}{\Delta E}\Big)
\Big(\frac{B}{10\rm T}\Big)^3\Big(\frac{10\mu\rm m}{w}\Big)^2 \Big(\frac{\mu}{3\times 10^5\rm cm^2/Vs}\Big)
\Big(\frac{I_x}{10^{-9}A}\Big)\Big(\frac{T}{m_a}\Big),
\end{equation}
where we have omitted the dependence of trivial factors such as $m_a$, $g_{\gamma}$ or $\rho_d$.
When a measurement of $\sigma_{xy}$ is performed in ten seconds, relevant number of electrons $\delta N_a^{\rm eff}$
excited to mobility gap
is $1.1\times 10^3$.

\vspace{0.1cm}
We compare it with the number of electrons thermally excited and determine which one is dominant for
the transition width $\Delta B$. 
Such a number $\delta N_t$ of electrons is given in the following.

\begin{equation}
\delta N_t=\int_{E_c-\delta}^{E_c+\Delta E}\frac{S\rho(E)dE}{1+\exp(\frac{E-E_f}{T})}\simeq
3.3\times 
10^7\exp(-\frac{m_a}{T})\Big(\frac{S}{10^{-3}\mbox{cm$^2$}}\Big)\Big(\frac{B}{10T}\Big)
\Big(\frac{0.5\times 10^{-4}\rm eV}{\Delta E}\Big)\Big(\frac{T}{m_a}\Big)
\end{equation}
with $E_f=E_c-\delta-m_a$, where we assume $\Delta E \gg m_a, \,\, \delta$,  and $m_a > T$. 

Therefore, we obtain the ratio of $\delta N_a^{\rm eff}\equiv \delta \dot{N}_a^{\rm eff}\times (10 \rm second)$ to $\delta N_t$,

\begin{equation}
\label{Nratio}
\frac{\delta N_a^{\rm eff}}{\delta N_t}\simeq 0.33\times 10^{-4}\exp(\frac{m_a}{T})
\Big(\frac{m_a}{10^{-5}\mbox{eV}}\Big)
\Big(\frac{B}{10T}\Big)^2\Big(\frac{10\mu\rm m}{w}\Big)^2\Big(\frac{\mu}{3\times 10^5\rm cm^2/Vs}\Big)
 \Big(\frac{I_x}{10^{-9}A}\Big)
\end{equation} 
where we have assumed that the time needed for measurement of $\sigma_{xy}$ at each strength of magnetic field $B$
is $10$ second.

\vspace{0.1cm}

\vspace{0.1cm}
Therefore, we have

\begin{equation}
\label{ratio}
\frac{\delta N_a^{\rm eff}}{\delta N_t}\sim 5.4 \quad \mbox{for} \quad \frac{m_a}{T}=12
\quad \mbox{but} \quad  \frac{\delta N_a^{\rm eff}}{\delta N_t}\sim 0.3 \quad \mbox{for} \quad \frac{m_a}{T}=9.
\end{equation}
\vspace{0.1cm}
The key point here is that this ratio does not depend on the sample area \( S \) or the width of the density of states \( \Delta E \). Except for the axion mass, all other parameters are determined by the experimental setup. Consequently, assuming an axion mass of \( 10^{-5} \, \rm eV \simeq 116 \, \rm mK \), the critical temperature obtained here is approximately \( 10 \, \rm mK \).  

Since rough approximations were used in the calculations, this critical temperature is not exact. However, the result suggests that the actual critical temperature likely falls within the range of \( 10 \, \rm mK \) to \( 30 \, \rm mK \).  
The value \( \delta N_a^{\rm eff}/\delta N_t \) is highly sensitive to temperature (\(\propto \exp(m_a / T)\)), meaning that the critical temperature is expected to stabilize within this range regardless of the approximations.

\vspace{0.1cm}

In the following section(\ref{f}), we determine that the axion mass is approximately \(10^{-5}\,\text{eV} \simeq 116\,\text{mK}\). Furthermore, analysis of experimental data 
( Fig.(\ref{6f}) and Fig.(\ref{6ff}) ) reveals that the critical temperature, below which axion dominance occurs, falls within the range of 20 mK to 30 mK. This result is consistent with our discussion.


\vspace{0.1cm}
Thus far, we have compared the number of electrons excited within the mobility gap by axion-induced microwaves with those excited by thermal effects to determine which is greater. To observe the actual impact of axion-induced microwaves, it is crucial to assess whether their effects are obscured by noise from blackbody radiation.

\vspace{0.1cm}
Thus, we calculate the signal-to-noise (SN) ratio associated with thermal noise. 
The thermal noise discussed here originates from blackbody radiation emitted by the sample.
To observe the effect of microwaves caused by axions, it is necessary for this SN ratio to be greater than $1$.

\vspace{0.1cm}
We take Fermi energy such as
$E_f+m_a =E_c-\delta+\delta E$. The absorption energy per unit time of axion microwave $P_a$ is given by
$\delta\dot{N}_a m_a$, where $\delta\dot{N}_a$ is evaluated at $T=0$ so that $\delta\dot{N}_a$ is proportional to $\delta E$.
The last factor $T/m_a$ in the formula $\delta\dot{N}_a$ in eq(\ref{8}) is replaced by $\delta E/m_a$. 
It is energy power $P_a$ absorbed by localized electrons
to transition to extended states in the energy range $E_c-\delta \sim E_c-\delta +\delta E$.

On the other hand, thermal noise $P_T$ associated with the energy $m_a\sim m_a+\delta E$ is given by

\begin{equation}
P_T=\frac{m_a\delta E}{2\pi(\exp(\frac{m_a}{T})-1)}\simeq \frac{m_a}{2\pi}\exp(-\frac{m_a}{T})\delta E
\end{equation}    
where $m_a> T$. Thus, SN ratio is given by

\begin{eqnarray}
\label{17}
&&\frac{P_a}{P_T/\sqrt{\delta E\times \rm 10s/2\pi}}\simeq \frac{\sqrt{2\pi}P_a\times \exp(\frac{m_a}{T})\sqrt{\delta E \times 10s}}{m_a\delta E} \nonumber \\ 
&\simeq& 2.6\times \Big(\frac{A}{10^4}\Big)^2 \Big(\frac{S}{10^{-3}\mbox{cm$^2$}}\Big)^2
\Big(\frac{0.5\times10^{-4}\mbox{eV}}{\Delta E}\Big)^2
\Big(\frac{\rho_d}{0.3\,\mbox{GeVcm$^{-3}$}}\Big)\Big(\frac{B}{10\rm T}\Big)^3
\Big(\frac{m_a}{10^{-5}\mbox{eV}}\Big)\exp\big(\frac{m_a}{10^{-5}\rm eV}\frac{20\rm mK}{T}\big)
\end{eqnarray}
where we have supposed that 
$\delta E=10^{-6}m_a$, $m_a=10^{-5}$eV and $10$ second for measuring $\sigma_{xy}$.
We have omitted a trivial factor $\frac{g_{\gamma}}{1.0}$.

It should be noted that the microwave frequency generated by the axion actually has a width of \( 10^{-6} (m_a/2\pi) \) around the central frequency \( \frac{m_a}{2\pi} \).  
This frequency width arises from the kinetic energy of axion dark matter, given by \( v_a^2 m_a / 2 \), where \( v_a \sim 10^{-3} \). Consequently, \( 10^{-6} m_a \) was adopted as \( \delta E \) in the calculation of the SN ratio.  
The result is consistent with our previous findings, confirming that, to observe the axion effect, a low temperature on the order of \( 10 \) mK is required.

\vspace{0.1cm}

The result depend on the sample's surface area $S$ and the width $\Delta E$ of its density of states. 
The larger the sample area, the easier it is to observe the effect of axions. In fact, as shown in the following chapters, axion effects are not observed in small samples. The influence of axions becomes detectable when the sample area \( S \) is at least \( 10^{-3} \,\text{cm}^2 \) or larger.

\vspace{0.1cm}
Indeed, the experiment shown in Fig.(\ref{6f}) uses a magnetic field of \( B \simeq 2.5 \) T and a Hall bar with dimensions \( S = 200 \, \mu\text{m} \times 800 \, \mu\text{m} \simeq 1.6 \times 10^{-3} \, \text{cm}^2 \). In this experiment, the axion effect was found to be dominant at \( T \sim 25 \) mK.  

On the other hand, the experiment shown in Fig.(\ref{10f}), where the axion effect is also dominant, uses a stronger magnetic field of \( B \simeq 3.4 \) T and a larger surface area of \( S = 6.6 \times 10^{-3} \, \text{cm}^2 \). Because the experiment in Fig.(\ref{10f}) employs a much stronger magnetic field and a larger sample area than the one in Fig. 7, the axion effect remains dominant over thermal effects even at a higher temperature of \( T = 35 \) mK.

%
%
%

We should make several comments for the observation of the axion effect. First, to observe it, we need large 
two dimensional surface area $S$ roughly larger than $10^{-3}\rm cm^2$. 
The absorption power $P_a=\delta \dot{N}_am_a$ of the axion radiation is proportional to $S^2$.
We also need the small width $\Delta E$ of the density of state $\rho \propto \sqrt{1 - ((E - E_{n\pm})/\Delta E)^2}$. 
The power is proportional to $(\Delta E)^{-2}$.  The smaller width is realized by larger mobility $\mu$ of sample.
The power is also larger as the magnetic field is stronger. It is
proportional to square of magnetic field $B^2$.
Finally, we need low temperature roughly less than $30$mK for axion effect to dominate thermal noise.

\section{examination of dependence of $\Delta B$ on temperature}

\subsection{Experiments with no axion effect}
Using the above results, we now examine several previous experiments, focusing on the temperature dependence of \(\Delta B\). Specifically, we analyze how \(\Delta B\) varies with temperature \(T\).

\begin{figure}[htp]
\centering
\includegraphics[width=0.65\hsize]{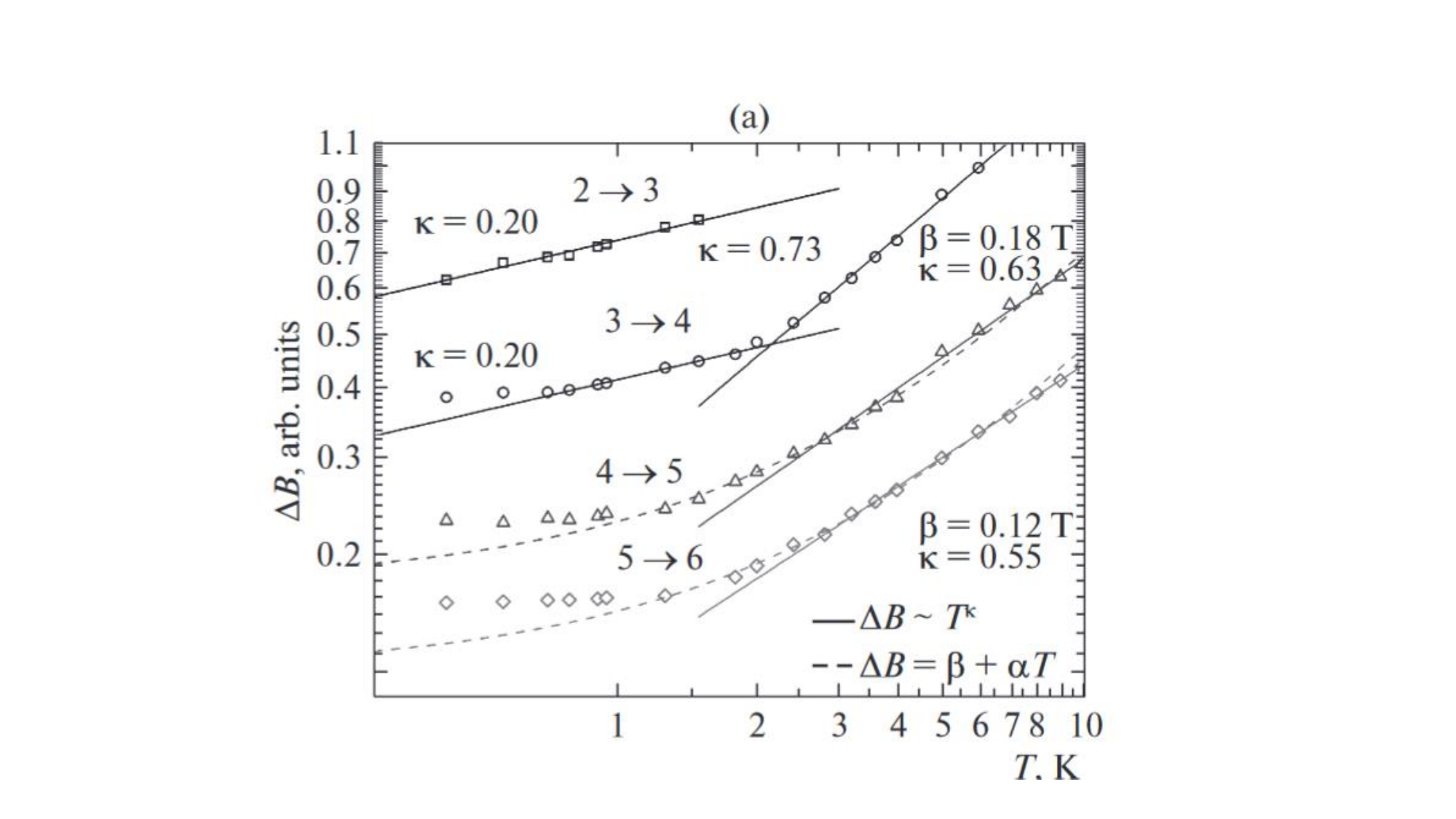}
\caption{$\Delta B$ decreases with decreasing $T$, but it saturates at saturation temperature $T=T_s\sim 1$K.
Saturated $\Delta B$ decreases as filling factor $\nu$ increases in transition $\nu \to \nu+1$ ( $\nu=2\sim 5$ ).
The size is $2.1\mu m\times 0.6\mu m$. Ref.\cite{exp1} }
\label{4f}
\end{figure}

First, we consider an experiment \cite{exp1} involving a small Hall bar with dimensions \(2.1 \, \mu\text{m} \times 0.6 \, \mu\text{m} \approx 1.3 \times 10^{-8} \, \text{cm}^2\) and mobility \(\mu = 2 \times 10^5 \, \text{cm}^2/\text{Vs}\). As shown in Fig.(\ref{4f}), \(\Delta B(T)\) for several plateau-plateau transitions saturates at a critical temperature \(T_c\). Here, the filling factor \(\nu = 2 \sim 6\) shown as e.g. $2\to 3$ in the figure
represents the ratio of the electron number density \(\rho_e\) to the Landau level degeneracy \(eB/2\pi\): \(\nu = 2\pi \rho_e/eB\).

Although the width \(\Delta B(T)\) in the figure differs slightly from our definition, it is essentially equivalent to the width we define. The observed saturation of \(\Delta B(T)\) with temperature is due to the saturation of the width \(\Delta E_f(T)\).

According to our analysis, saturation with temperature would not normally occur, as the width \(\Delta E_f = 2\delta(T) + 4T\) does not saturate at any temperature \(T\) in the absence of an axion effect. However, the figure shows that \(\Delta B\) saturates at a temperature \(T_s \sim 1\) K, particularly for the transitions \(\nu = 3 \to 4\), \(\nu = 4 \to 5\), and \(\nu = 5 \to 6\). This observed saturation can be explained under the condition \(2\delta(T) \gg 4T\).


\begin{equation}
\Delta E_f=2\delta(T)+4T \sim 2\delta(T) \quad \mbox{for}  \quad T \le O(1)\rm K . 
\end{equation}
We thus find that the width \(\Delta E_f\) saturates at the temperature \(T_s = T_c \sim 1\) K because the mobility gap \(\delta(T)\) reaches a constant value at the critical temperature \(T = T_c\). This in turn leads to the saturation of the width \(\Delta B\). The critical temperature \(T_c\) is defined such that \(\delta(T \leq T_c) = \delta(T = 0)\). Without the assumption \(2\delta(T) \gg 4T\), the width would not saturate. Given the very small size of the Hall bar, it is plausible that the mobility gap \(\delta\) is large enough for this assumption, \(2\delta(T) \gg 4T\), to hold, especially around \(T \sim 1\) K. In any case, the observed saturation of \(\Delta B\) with temperature \(T\) supports the validity of this assumption.

\vspace{0.1cm}

We also observe that the saturated value of \(\Delta B\) decreases as the filling factor \(\nu\) increases in the transition \(\nu \to \nu + 1\) (\(\nu = 2 \sim 5\)). This suggests that the mobility gap \(\delta\) decreases with increasing \(\nu\). 
Since the wave functions in higher Landau levels extend further than those in lower levels, they are more influenced by the disorder potential, resulting in a reduction in the spatial extent of their energy eigenstates. In other words, the localization length of wave functions in higher Landau levels is shorter than in lower levels. It leads to
a smaller mobility gap in the density of states in higher Landau levels. This is a general feature of mobility gap \(\delta\).


\vspace{0.1cm}

\begin{figure}[htp]
\centering
\includegraphics[width=0.84\hsize]{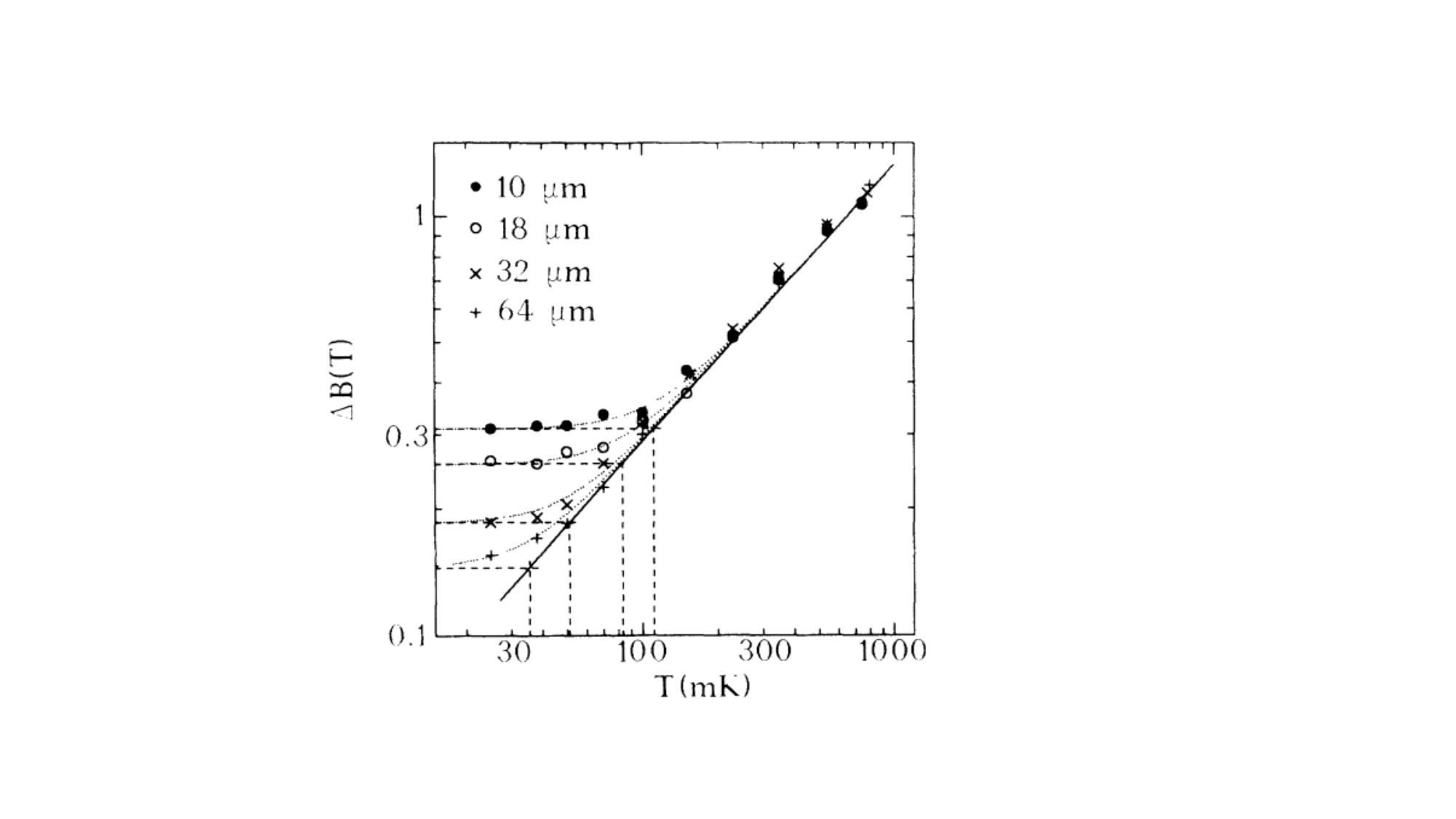}
\caption{Saturation temperature $T_s ( <50$mK ) decreases as the width $W$ increases, 
ranging from $10\mu m$ to $64 \mu m$. Ref.\cite{exp2} }
\label{5f}
\end{figure}

We proceed to examine an experiment presented in Fig.(\ref{5f}) from \cite{exp2}, which used Hall bars larger than the previous case, with widths \(W = 10 \, \mu \text{m},\, 18 \, \mu \text{m},\, 32 \, \mu \text{m},\,\) and \(64 \, \mu \text{m}\), with a width to length ratio of 1:3. For example, the surface area of a \(32 \, \mu \text{m} \times 96 \, \mu \text{m}\) sample is about \(3 \times 10^{-5} \, \text{cm}^2\), and the mobility is \(\mu = 1.55 \times 10^5 \, \text{cm}^2/\text{Vs}\). The sample consists of \(\text{Al}_x\text{Ga}_{1-x}\text{As}/\text{Al}_{0.32}\text{Ga}_{0.68}\text{As}\) with \(x = 0.33\). The data reveal that \(\Delta B(T)\) saturates at temperatures below approximately \(T_s \sim 50\) mK and decreases with increasing width \(W\). Notably, for the largest sample (\(64 \, \mu \text{m} \times 192 \, \mu \text{m}\)), \(\Delta B\) does not saturate even at the lowest experimental temperature of \(T = 25\) mK. The saturation temperature \(< 50\) mK also decreases as the sample size exceeds \(10^{-5} \, \text{cm}^2\).

The sample sizes in this experiment are still too small for any axion contribution to appear. Thus, the saturation observed here suggests that the condition \(2\delta(T) \gg 4T\) remains valid, leading to \(\Delta E_f \simeq 2\delta(T)\) saturating at the critical temperature associated with the mobility gap, which is approximately \(T_c \sim 50\) mK for \(W = 10 \, \mu \text{m}\), \(T_c \sim 40\) mK for \(W = 18 \, \mu \text{m}\), and \(T_c \sim 25\) mK for \(W = 32 \, \mu \text{m}\). 

In the largest sample (\(64 \, \mu \text{m} \times 192 \, \mu \text{m}\)), it is likely that \(\delta(T)\) is small enough that the assumption \(2\delta(T) \gg 4T\) does not hold.

Although \(2\delta(T) \gg 4T\) at low temperatures (\(T \le 50\) mK), we observe that \(2\delta(T) \ll 4T\) at higher temperatures (\(T \ge 200\) mK), as all curves converge at \(T \ge 200\) mK. This convergence suggests that \(\delta(T)/T\) increases rapidly as \(T\) decreases.

\vspace{0.1cm}

Furthermore, it is evident that the mobility gap \(\delta(T)\) decreases with increasing width \(W\). This trend aligns with the general properties of the mobility gap, as discussed in the previous section.

\vspace{0.1cm}

As the width \( W \) of the Hall bar increases, the critical temperature \( T_c(W) \) decreases, since a lower temperature is sufficient for the phase coherence length to match the larger size of the Hall bar. In fact, it has been shown experimentally \cite{exp3} that \( T_c(W) \propto 1/W \) for Hall bars with widths ranging from \( W = 100 \, \mu \text{m} \) to \( 1000 \, \mu \text{m} \) and a width to length ratio of \( 2.5:4.5 \). This relationship \( T_c(W) \propto 1/W \) seems to approximately hold in the experiment shown in Fig.(\ref{5f}).

\vspace{0.1cm}

It is worth noting that the observed relationship \( T_c(W) \propto 1/W \) applies specifically to samples with short-range disorder potentials; for example, the sample is composed of \(\text{Al}_x\text{Ga}_{1-x}\text{As/Al}_{0.32}\text{Ga}_{0.68}\text{As} \) with \( x = 0.85 \), similar to the sample used in Fig.(\ref{5f}). The sample’s high mobility, \(\mu = 8.9 \times 10^5 \, \text{cm}^2/\text{Vs}\), results in a relatively large mobility gap \(\delta\), so that \(\Delta E_f = 2\delta(T) + 4T \approx 2\delta(T)\). The dependence of \( T_c(W) \) on \( W \) is thus influenced by the nature of the disorder potential. Although the exact formula \( T_c(W) \propto 1/W \) may not hold universally for all disorder potentials, the trend \( T_c(W) \to 0 \) as \( W \to \infty \) generally persists.

It may appear that the axion effect could manifest in this experiment \cite{exp3}, 
given the sample’s large size (e.g., \( 500 \, \mu \text{m} \times 900 \, \mu \text{m} = 4.5 \times 10^{-3} \, \text{cm}^2 \)). However, the applied magnetic field is relatively weak at \( B = 1.4 \, \text{T} \), so the axion effect remains negligible in this context.

%
%
%
%

\subsection{Experiments with axion effect}

Up to now, the axion contribution has been negligibly small, meaning the width \(\Delta E_f\) is given by \(\Delta E_f = 2\delta(T) + 4T\). As we have shown, the critical temperature \(T_c\) matches the saturation temperature \(T_s\), where \(\Delta B\) (or \(\Delta E_f\)) stabilizes, as \(\Delta E_f = 2\delta(T) + 4T \approx 2\delta(T) = 2\delta(T_c)\) for \(T \le T_c\). Thus, the saturation temperature \(T_s\) depends on the Hall bar’s size, a phenomenon referred to as the finite-size effect in prior research. This is the case when the axion contribution remains negligible.

On the other hand, when the axion contribution is significant, \(\Delta E_f\) is not given by \(2\delta(T) + 4T\). Instead, \(\Delta E_f = 2\delta(T) + 4m_a\) for \(T < m_a\), while \(\Delta E_f = 2\delta(T) + 4T\) for \(T > m_a\). This axion contribution becomes comparable to the thermal component at very low temperatures and in large Hall bar sizes. In this regime, the saturation temperature \(T_s = m_a\) is not necessarily dependent on the Hall bar’s size. ( We should remember that actual saturation temperature is less than $m_a$ as shown in
the previous section(\ref{axion}).
We tentatively use the equality $T_s=m_a$ in theoretical discussion
in order to show that axion effect becomes manifest and dominant at low temperature. ) 

\vspace{0.1cm}

\begin{figure}[htp]
\centering
\includegraphics[width=0.67\hsize]{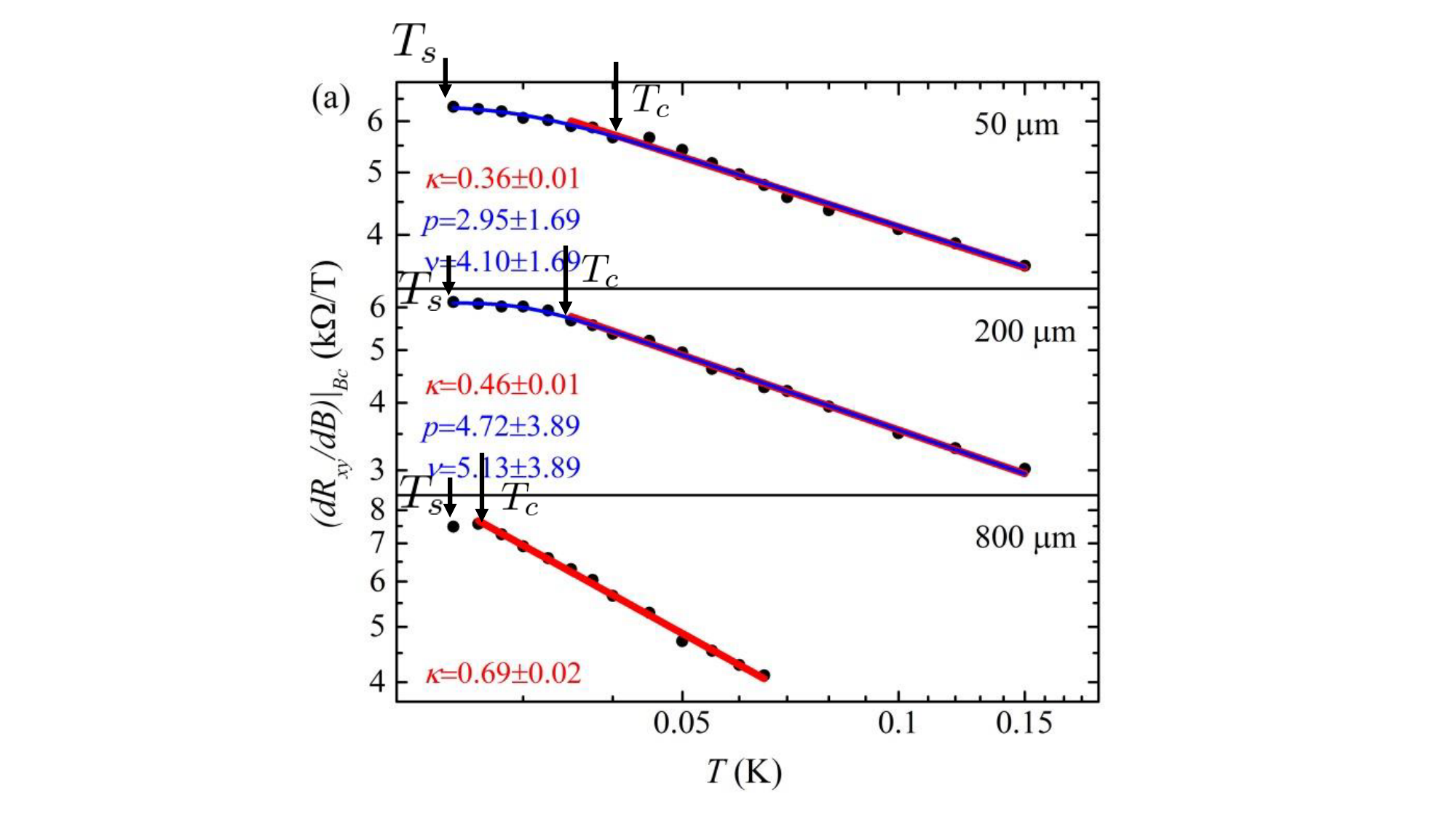}
\caption{$R_{xy}$ in the figure is identical to $\rho_{xy}$. We see that saturation temperatures $\sim 25$mK for all samples.
Sizes of Hall bars are $50\mu$m$\times 200\mu$m, $200\mu$m$\times 800\mu$m 
and $800\mu$m$\times 1600\mu$m.  We have indicated saturation temperature $T_s$ and critical temperature $T_c$ by allows. 
Ref.\cite{exp4} }
\label{6f}
\end{figure}

An intriguing experiment \cite{exp4}, shown in Fig.(\ref{6f}), reveals that the saturation temperature \(T_s \approx 20\) mK remains constant even as the Hall bar’s size changes substantially. In this experiment, the width \(W\) varies from \(50 \, \mu \text{m}\) to \(800 \, \mu \text{m}\) with a width to length ratio of \(1:4\), leading to a surface area of, for example, \(200 \, \mu \text{m} \times 800 \, \mu \text{m} = 1.6 \times 10^{-3} \, \text{cm}^2\). The sample’s high mobility, \(\mu = 2.8 \times 10^5 \, \text{cm}^2/\text{Vs}\), and composition (\(\text{GaAs/Al}_{0.32}\text{Ga}_{0.68}\text{As}\) with no short-range disorder potential) imply a relatively narrow extension \(\Delta E\) in the density of states compared to samples with short-range disorder. The sample size is large enough to allow the axion contribution to emerge, strongly suggesting its presence. (As discussed, the temperature dependence of \(dR_{xy}/dB\propto 1/\Delta B\) effectively reflects the temperature dependence of the width \(\Delta B\).)

\vspace{0.1cm}

This behavior can be understood as follows: when \(T \gg m_a\), \(\Delta E_f = 2\delta(T) + 4T\). As \(T\) decreases, it eventually reaches the critical temperature \(T = T_c\), where \(\delta(T < T_c) = \delta(T_c)\). However, even at this point, \(\Delta E_f = 2\delta(T_c) + 4T\) continues to decrease with falling \(T\). Finally, when \(T\) reaches \(m_a\), the width \(\Delta E_f = 2\delta(T_c) + 4m_a\) saturates, as electron distribution becomes dominantly influenced by microwave frequency \(m_a / 2\pi\) associated with the axion. Consequently, the saturation temperature \(T_s = m_a\) becomes independent of the Hall bar’s size, aligning well with experimental results in Fig.(\ref{6f}).

\vspace{0.1cm}
We should note that
when we closely look the figure, we find that saturation temperature is slightly larger as the size of Hall bar is larger.
This is consistent with the equation(\ref{17}).

\vspace{0.1cm}

The experimental authors attribute this size independence to “intrinsic decoherence,” though its source is not clear. In contrast, we suggest that this decoherence is induced by axion dark matter, an extrinsic rather than intrinsic effect.

\begin{figure}[htp]
\centering
\includegraphics[width=0.77\hsize]{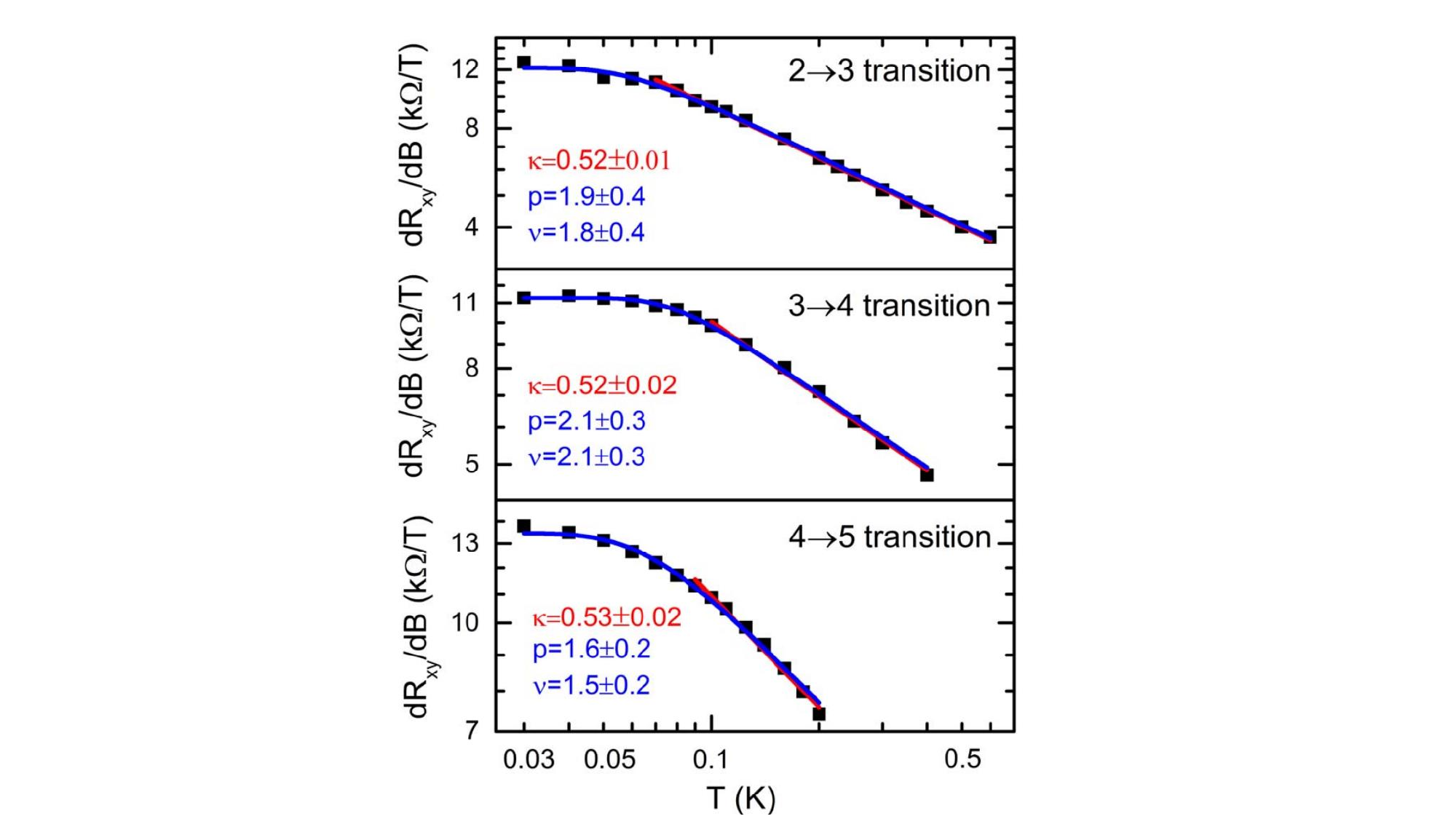}
\caption{$R_{xy}$ in the figure is identical to $\rho_{xy}$.
The saturated values $d R_{xy}/dB$ are almost independent on 
the transitions, $\nu=2\to 3$, $\nu=3\to 4$ and $\nu=4\to 5$.  
We see that saturation temperatures is roughly $\sim 30$mK for all
transitions, which is identical to the saturation temperature $25$mK shown in Fig.(\ref{6f}).  
Size of Hall bars is $150\rm \mu m\times 1320\mu m\simeq 2\times 10^{-3}cm^2$. 
It is formed of $\rm GaAs/Al_{0.32}Ga_{0.68}As$. The mobility $\mu=2.9\times \rm cm^2/Vs$.
Ref.\cite{exp44} }
\label{6ff}
\end{figure}

\vspace{0.1cm}
%
Additionally, an experiment \cite{exp44} using a large \(\text{GaAs/Al}_{0.32}\text{Ga}_{0.68}\text{As}\) sample with dimensions \(150 \, \mu\text{m} \times 1320 \, \mu\text{m} \approx 2 \times 10^{-3} \, \text{cm}^2\) has shown that the saturated values of \(d\rho_{xy}/dB\) are nearly independent across different plateau transitions (\(\nu = 2 \to 3\), \(\nu = 3 \to 4\), and \(\nu = 4 \to 5\)), as illustrated in Fig.(\ref{6ff}). By noting the formula $d\rho_{xy}/dB \approx \Delta \rho_{xy} / \Delta B$ with $\Delta \rho_{xy} = 2\pi / e^2 \times 1 / (n(n+1))$ with $n=2\sim 4$, 
this observation suggests that $\Delta E_f\simeq \delta(T_s)$, namely 
the mobility gap \(\delta(T_s)\) diminishes as $n$ increases as discussed above.
Furthermore, the saturation temperatures \(T_s\) for each transition are approximately \(T_s \sim 30\) mK.

Notably, this saturation temperature \(T_s \sim 30\) mK is very close to the \(T_s \approx 20\) mK observed in another experiment \cite{exp4}. Such a nearly identical saturation temperature across different samples strongly points to the validity of our
previous discussion about the saturation temperature below which the axion effect dominates the thermal effect.

\vspace{0.1cm}
In Fig.(\ref{6f}), we illustrate the presence of critical temperatures \( T_c \), where the curves begin to show slight bending. The values of \( T_c \) were selected visually and are indicated by arrows. The critical temperature \( T_c \) decreases with increasing Hall bar size, as shown, but does not follow the \( T_c(W) \propto 1/W \) relation. Instead, the dependence of \( T_c \) on size is weaker here compared to the expected \( T_c(W) \propto 1/W \) behavior. This discrepancy likely arises from differences in semiconductor composition between the samples. The first sample, which follows the \( T_c(W) \propto 1/W \) relation, is based on \(\text{Al}_x\text{Ga}_{1-x}\text{As}/\text{Al}_{0.32}\text{Ga}_{0.68}\text{As}\) with \( x = 0.85 \), whereas the second sample is a \(\text{GaAs}/\text{AlGaAs}\) heterostructure. The high aluminum content (\( x = 0.85 \)) in the first sample introduces short-range disorder potential \( V_s \) alongside the existing long-range potential \( V_l \). By contrast, in the \(\text{GaAs}/\text{AlGaAs}\) sample, without such contamination (\( x = 0 \)), only the long-range potential \( V_l \) is present. The additional short-range potential in the contaminated sample likely results in the observed difference in \( T_c(W) \) dependence on \( W \). Generally, short-range disorder potential has a stronger effect than long-range potential, leading to a larger density-of-states extension \(\Delta E_s\) in the presence of \( V_s \) compared to \(\Delta E_l\), where only \( V_l \) is present (\( \Delta E_l < \Delta E_s \)). According to Eq.(\ref{17}), this can make the axion effect more readily detectable in samples with only long-range potential, even for smaller sample sizes.

Thus, the experiments indicate that the axion effect appears because the saturation temperature \( T_s = m_a \) is almost independent of the Hall bar size and sample composition. This strongly supports the existence of the axion. 
Additionally, almost identical values of \( dR_{xy}/dB \) (or equivalently, \( \delta(T_s, m_a/2\pi) \)) across varying Hall bar sizes and plateau transitions may also be a signature of the axion’s presence.

\vspace{0.1cm}

We would like to clarify a point. As previously mentioned, the microwave frequency \( f \) does not correspond to an actual frequency in the electron distribution. It is the cut-off frequency assumed tentatively in the distribution. 
This frequency appears in the expression \( \Delta E_f = 2\delta + 8\pi f \). Therefore, 
even if we observe the relation \( T_s = 2\pi f_s = m_a \sim 25 \, \text{mK} \), this does not necessarily imply an axion mass of \( m_a = 25 \, \text{mK} \sim 2.2 \times 10^{-6} \, \text{eV} \). As we have explained in previous section (\ref{axion}), the saturation temperature
below which the axion effect dominates thermal one,
is in general lower than the axion mass $m_a$.


\section{examination of dependence of $\Delta B$ on frequency $f$}
\label{f}
Now, let us consider prior experiments involving the application of externally imposed microwaves. In these experiments, the width \(\Delta B\) corresponds to the width of the peak in the longitudinal electric conductance, \(\sigma_{xx}\). Although the DC conductance vanishes within plateaus, it rises to a non-zero value during plateau-plateau transitions, forming a peak. This peak width is analogous to the \(\Delta B\) defined earlier in this paper. By measuring the absorption power in two-dimensional electron systems, the real part of the electric conductivity, \(\text{Re}(\sigma_{xx})\), can be obtained under microwave irradiation \cite{exp9,exp10}. In the measurements discussed below, \(\Delta B\) denotes the peak width of \(\text{Re}(\sigma_{xx})\). Since we are concerned only with the variation in width relative to microwave frequency, the precise definition of width is not critical in this context.


\subsection{Experiments with no axion effect}
\begin{figure}[htp]
\centering
\includegraphics[width=0.75\hsize]{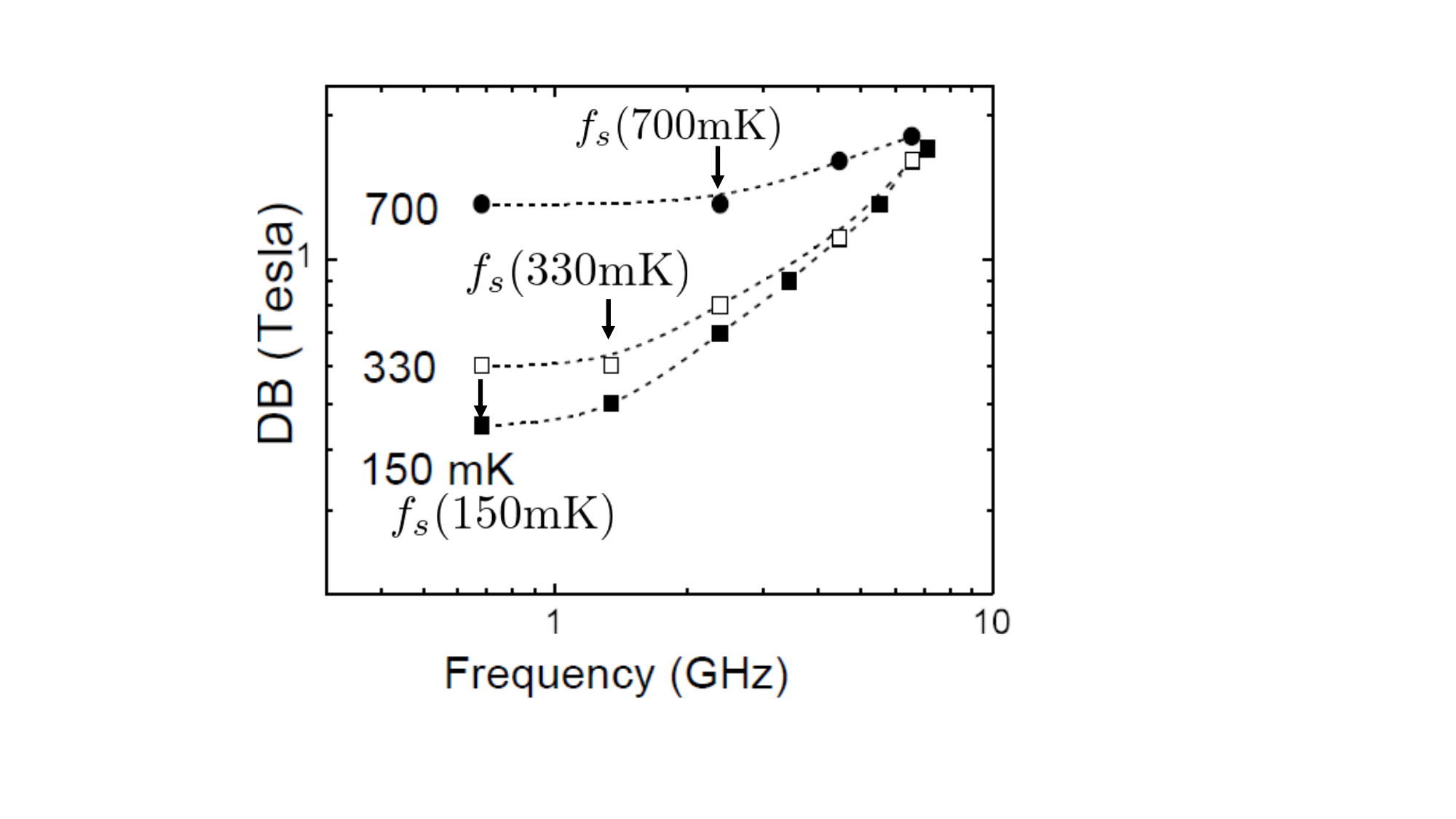}
\caption{The dependence of \( \Delta B \) on frequency \( f \) is shown for three different temperatures in a small sample measuring \( 164 \, \mu\text{m} \times 64 \, \mu\text{m} \). As the frequency decreases, \( \Delta B \) also decreases, but it eventually saturates at \( f = f_s(T) \). The saturation frequency \( f_s(T) \) decreases as the temperature \( T \) is lowered. The arrows indicate the saturation frequencies \( f_s \), demonstrating that \( f_s(150 \, \text{mK}) < 1 \, \text{GHz} \) even at the relatively high temperature of \( T = 150 \, \text{mK} \). Ref.\cite{exp5}}
\label{7f}
\end{figure}

First, we examine the experiment \cite{exp5} conducted with a relatively small sample of size \( 164 \mu m \times 64 \mu m \sim 10^{-4} \, \text{cm}^2 \). Given the small size of the sample, we expect the axion contribution to be negligible at higher temperatures, such as \( T > 100 \, \text{mK} \). The behavior of the width \(\Delta B\) as a function of microwave frequency is shown in Fig. (\ref{7f}).

The mobility of the sample is relatively low, with \( \mu = 3.4 \times 10^4 \, \text{cm}^2/\text{Vs} \). The sample, composed of \( \rm GaAs/AlGaAs \), is smaller in size compared to others discussed later. The width \( \Delta E_f \) follows the relation \( \Delta E_f = 2\delta(T,f) + 8\pi f \) for higher frequencies (\( f > T/2\pi \)). As the frequency decreases, the mobility gap \( \delta(T,f) \) first saturates at \( f = f_c(T) > T/2\pi \), where \( \delta(T,f) = \delta(T,f_c(T)) \) for \( f \le f_c(T) \). Therefore, the width is given by \( \Delta E_f = 2\delta(T,f_c(T)) + 8\pi f \). As the frequency decreases further, \( \Delta E_f \) continues to decrease until it saturates at the frequency \( f_s(T) = T/2\pi \), where the width becomes \( \Delta E_f = 2\delta(T,f_c(T)) + 4T \). This represents a transition from frequency dominance to temperature dominance in the electron distribution at \( f_s(T) = T/2\pi \).

\vspace{0.1cm}

This interpretation aligns with the experimental observations, such as \( f_s(700 \, \text{mK}) \sim 2f_s(330 \, \text{mK}) \) and \( f_s(330 \, \text{mK}) \sim 2f_s(150 \, \text{mK}) \), which suggests that \( f_s(T) \propto T \). In this context, we assume a large critical frequency \( f_c(T) > T/2\pi \) due to the small size of the Hall bar, which leads to a large mobility gap \( \delta(f) \) and a corresponding high critical frequency \( f_c \). For very large Hall bars, however, the critical frequency \( f_c(T) \) would satisfy \( f_c(T) < T/2\pi \), as discussed later. This framework allows us to understand the experiment conducted at temperatures \( T \ge 150 \, \text{mK} \), where the axion effect can be neglected.

\vspace{0.1cm}

It is important to note that in Fig.(\ref{7f}), the curves converge at a frequency of approximately 8 GHz. This indicates that for frequencies \( f \ge 8 \, \text{GHz} \) and temperatures \( T \le 700 \, \text{mK} \), the condition \( 2\delta(T,f) \ll 8\pi f \) holds. However, as the frequency decreases below 8 GHz, \( 2\delta(T,f)/8\pi f \) rapidly increases 
and becomes of the order of $1$.

\vspace{0.1cm}

\begin{figure}[htp]
\centering
\includegraphics[width=0.73\hsize]{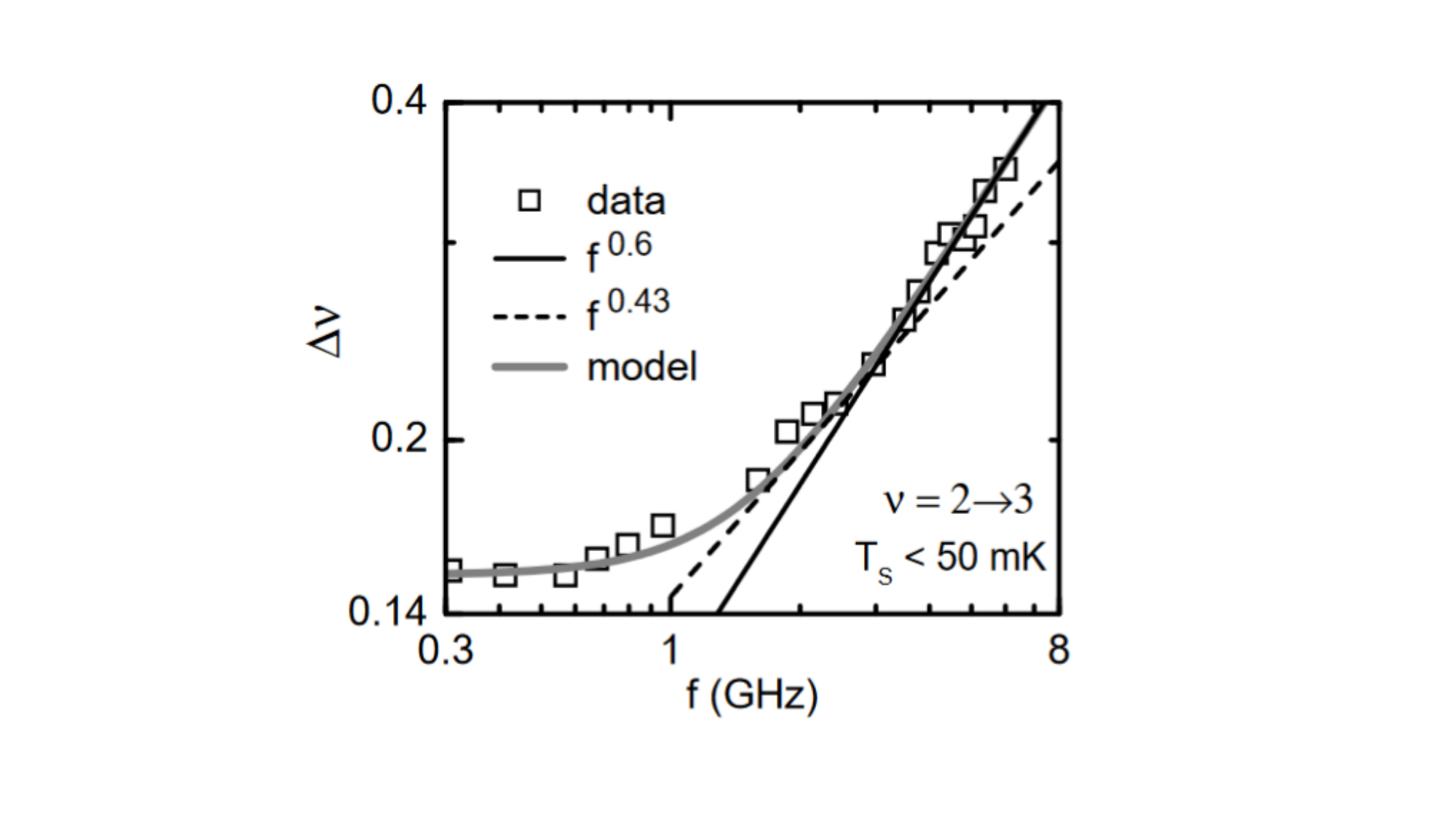}
\caption{The dependence of \( \delta\nu \) on frequency at a low temperature of \( 50 \, \text{mK} \) is shown for a sample with Corbino geometry, having a radius of approximately \( 800 \, \mu\text{m} \) and a width of \( 20 \, \mu\text{m} \). The behavior of \( \delta\nu = -\Delta B / \nu \) closely mirrors that of \( \Delta B \). The results indicate a saturation frequency of \( f_s(50 \, \text{mK}) \approx 0.6 \, \text{GHz} \). Ref.\cite{exp6} }
\label{8f}
\end{figure}

Next, we analyze the experiment \cite{exp6} conducted at temperatures below 50 mK, shown in Fig.(\ref{8f}). The sample, also made of \( \rm GaAs/AlGaAs \), has a mobility of \( \mu = 3.5 \times 10^5 \, \text{cm}^2/\text{Vs} \). The Hall bar in this experiment follows a Corbino geometry, where two-dimensional electrons are confined between an outer circle with a radius of 820 µm and an inner circle with a radius of 800 µm, giving a surface area of approximately \( 10^{-3} \, \text{cm}^2 \).

At such low temperatures, the axion effect could potentially appear, but it seems not to manifest in this case. The surface area of the Hall bar is still not large enough for the axion-induced radiation to absorb a significant amount of energy. In the figure, the width of the filling factor \( \Delta \nu \equiv \Delta (2\pi \rho_e / eB) = -\Delta B \nu \) is plotted. The behavior of \( \Delta \nu \) as a function of frequency \( f \) is identical to the behavior of \( \Delta B \) with respect to \( f \).

The saturation frequency at \( T = 50 \, \text{mK} \) is found to be \( f_s(50 \, \text{mK}) \sim 0.6 \, \text{GHz} \). This result is consistent with expectations from earlier experiments, where the saturation frequency decreases as the temperature is lowered. However, given that a different Hall bar geometry is used here, a direct comparison requires careful consideration. To further illustrate this, we present an additional experiment \cite{hohls2} that uses the same geometrical configuration (Corbino geometry) and a \( \rm GaAs/AlGaAs \) sample of identical size, but at a higher temperature of 300 mK.

\begin{figure}[htp]
\centering
\includegraphics[width=0.8\hsize]{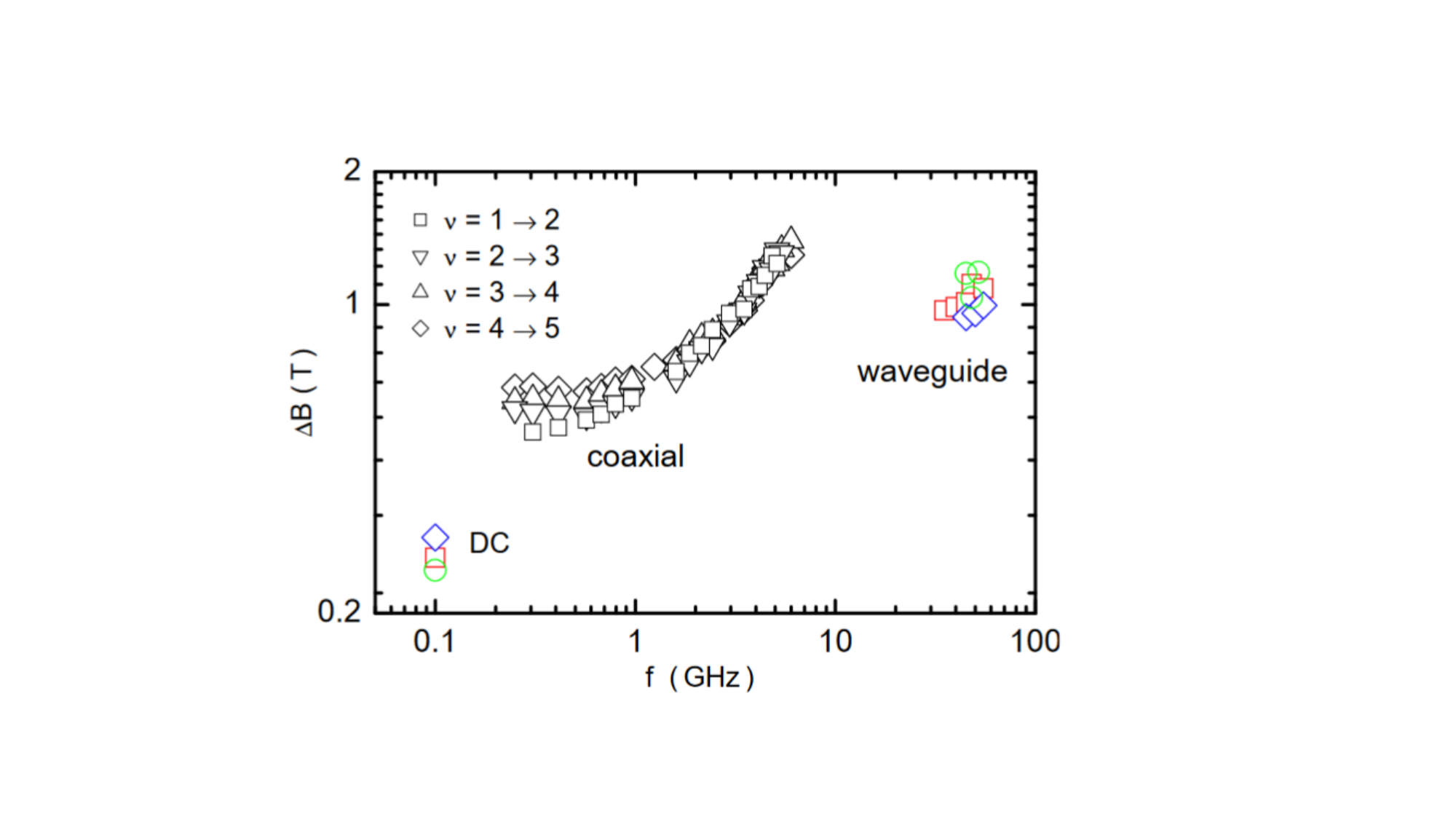}
\caption{The dependence of \( \Delta B \) on microwave frequency at a temperature of \( 300 \, \text{mK} \) is shown for the same sample as in Fig. \ref{8f}. The plot displays various \( \Delta B \) values corresponding to different plateau-plateau transitions characterized by \(\nu\). For the transition \(\nu = 2 \rightarrow 3\), a saturation frequency of \( f_s(300 \, \text{mK}) \approx 0.9 \, \text{GHz} \) is observed.  Ref.\cite{hohls2} }
\label{9f}
\end{figure}

In Fig.(\ref{9f}), we present the results from the experiment at \( T = 300 \, \text{mK} \) \cite{hohls2}. 
The curve ($\nu=2\to 3$) labeled "coaxial" corresponds to the transition \( \nu = 2 \to 3 \) and matches the result shown in Fig.(\ref{8f}). We observe that the saturation frequency at \( T = 300 \, \text{mK} \) is \( f_s(300 \, \text{mK}) \sim 0.9 \, \text{GHz} \), which can be compared with the saturation frequency at \( T = 50 \, \text{mK} \), \( f_s(50 \, \text{mK}) \sim 0.6 \, \text{GHz} \).
  
%
%

The results indicate that while the saturation frequency \( f_s(T) \) decreases with temperature \( T \), the relationship \( f_s(T) \propto T \) does not hold. This sample, being significantly larger than the one used in the experiment shown in Fig.(\ref{7f}), has a smaller critical frequency \( f_c(T) \). It is likely that \( f_c(T) < T / 2\pi \). Therefore, the behavior of the width \( \Delta E_f \) follows the pattern: when the frequency \( f \) is large (\( f > T / 2\pi \)), \( \Delta E_f = 2\delta(T,f) + 8\pi f \). As the frequency decreases, it first reaches \( f = T / 2\pi \) (assuming \( f_c(T) < f \)). At this point, frequency dominance is replaced by temperature dominance, and the width becomes \( \Delta E_f = 2\delta(T,f) + 4T \). Further decreases in frequency cause the width to decrease, but when \( f = f_c(T) \), the width saturates at \( \Delta E_f = 2\delta(T,f_c) + 4T \). Thus, the saturation frequency is \( f_s(T) = f_c(T) \), which is not necessarily proportional to temperature. This interpretation aligns with and is consistent with the previous experimental results, where the axion contribution remains negligible.

\vspace{0.1cm}
It is worth noting that even for small samples, such as \( 164 \, \mu m \times 64 \, \mu m \), or at relatively high temperatures like \( 150 \, \text{mK} \), the saturation frequency is still below 1 GHz. In general, the saturation frequency tends to decrease as the sample size increases (or temperature decreases), a behavior observed in the absence of an axion effect. Consequently, we expect that in experiments involving larger samples and lower temperatures than those presented here, the saturation frequency would be much lower than 1 GHz if the axion effect is negligible. As shown in Fig.(\ref{11f}), for an example where the sample size is large (\( 14 \, \text{mm} \times 30 \, \mu m \sim 4 \times 10^{-3} \, \text{cm}^2 \)) and the temperature is low (\( 50 \, \text{mK} \)), 
but the low mobility \(\mu = 4 \times 10^4 \ \text{cm}^2/\text{Vs}\), 
the saturation frequency is also less than 1 GHz. The authors in reference \cite{hohls2} report that saturation frequencies are below 1 GHz at around 100 mK in various experiments conducted up to the time of the paper's publication.

\subsection{Experiments with axion effect}

%

\begin{figure}[htp]
\centering
\includegraphics[width=0.65\hsize]{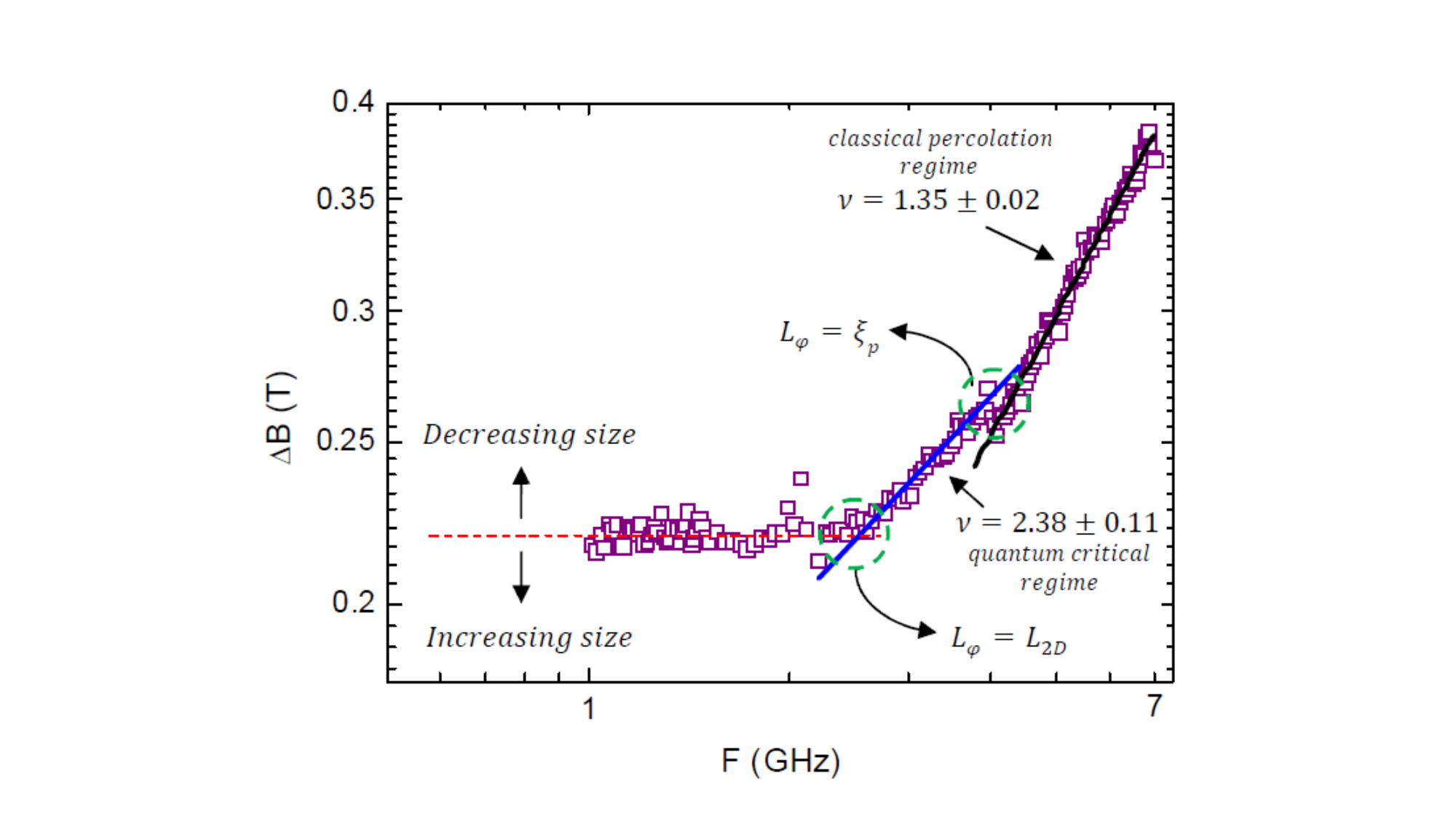}
\caption{The observed saturation frequency of \( f_s \simeq 2.4 \, \text{GHz} \) at \( T = 35 \, \text{mK} \) corresponds to an axion mass of \( m_a \simeq 10^{-5} \, \text{eV} \). This result was obtained for a sample size of \( 22 \, \text{mm} \times 30 \, \mu\text{m} \). A similar saturation frequency \( f_s \) was also observed in a smaller sample measuring \( 5.5 \, \text{mm} \times 30 \, \mu\text{m} \). Ref.\cite{exp9} }
\label{10f}
\end{figure}

Up to this point, we have examined relatively small Hall bars that do not absorb a sufficiently large amount of microwave radiation generated by the axion to produce an observable effect. We now turn our attention to an experiment conducted using a larger Hall bar with dimensions \( 22 \, \text{mm} \times 30 \, \mu m \), corresponding to a surface area of \( 6.6 \times 10^{-3} \, \text{cm}^2 \). The sample is composed of \( \rm GaAs/AlGaAs \), with a mobility of \( \mu = 3.8 \times 10^5 \, \text{cm}^2/\text{Vs} \). External microwaves are applied via a coplanar waveguide structure. The results of the experiment at a low temperature of \( 35 \, \text{mK} \) are shown in Fig. (\ref{10f}). Unlike the previous results, the saturation frequency is much higher, with \( f_s(35 \, \text{mK}) \approx 2.4 \, \text{GHz} \). It is noteworthy that this experiment achieves significantly higher frequency resolution than those conducted previously.

\vspace{0.1cm}
Based on the previous analysis, we would have expected the saturation frequency to be much lower than 1 GHz, given that the temperature is lower and the sample size is larger than in the earlier experiments. The larger size leads to a lower critical frequency \( f_c \) compared to the smaller samples. However, the result here is quite unexpected.

\vspace{0.1cm}

The observed result can only be understood by considering the axion contribution. In this case, the axion effect becomes observable in the large sample at a low temperature of \( 35 \, \text{mK} \). The axion dark matter generates microwaves with a frequency of \( m_a/2\pi \), which is always present, even when external microwaves with various frequencies \( f \) are applied. 

At large frequencies, \( f > m_a/2\pi \), the width is given by \( \Delta E_f = 2\delta(T = 35 \, \text{mK}, f) + 8\pi f \). As the frequency decreases, the width also decreases. However, when the frequency reaches \( f = m_a/2\pi \), the dominance shifts from the external microwave to the axion-generated microwave. This means that for \( f < m_a/2\pi \), the electron distribution is primarily influenced by the microwave generated by the axion, while for \( f > m_a/2\pi \), it is determined by the external microwave.

The axion dominance also implies that the mobility gap \( \delta(T, f) \) is determined by the axion-generated microwave, i.e., \( \delta(T, f = m_a/2\pi) \). In other words, electrons in localized states can transition to extended states by absorbing the higher energy \( m_a \) associated with the axion, but they cannot hop to extended states by absorbing an energy \( 2\pi f \) unless \( f \) exceeds \( m_a/2\pi \). Therefore, \( \delta(T, m_a) > \delta(T, f) \) when \( m_a > 2\pi f \).

Hence, the width \( \Delta E_f \) is given by
 $\Delta E_f = 2\delta(T = 35 \, \text{mK}, m_a/2\pi) + 4m_a$,
when the frequency is below the saturation frequency \( f_s = m_a/2\pi \), which is determined by the axion mass \( m_a \). The experimental results suggest the presence of an axion with a mass \( m_a \simeq 10^{-5} \, \text{eV} \), as indicated by the saturation frequency of approximately \( 2.4 \, \text{GHz} \).

\begin{figure}[htp]
\centering
\includegraphics[width=0.65\hsize]{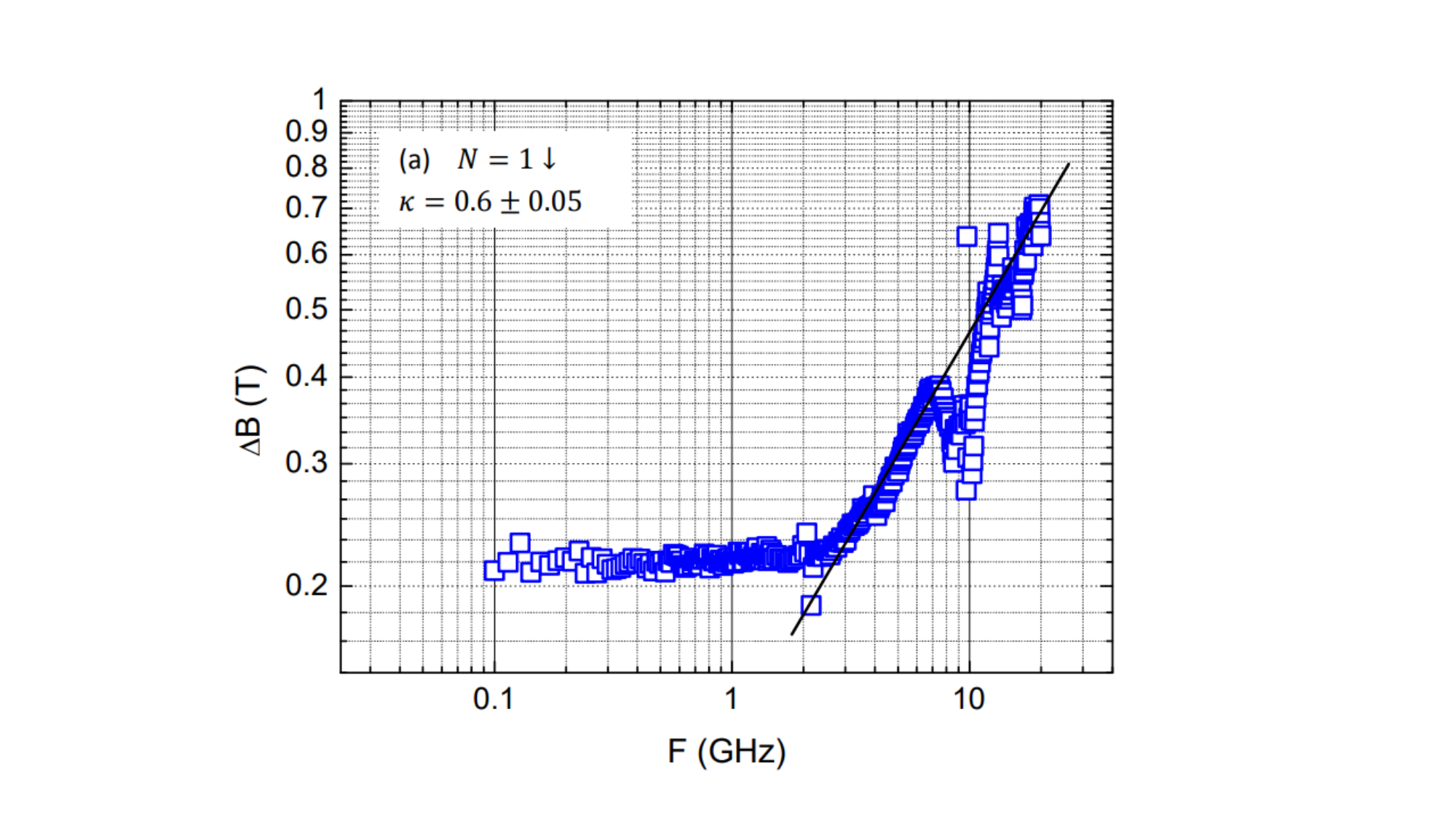}
\caption{A saturation frequency of \( f_s \sim 2.4 \, \text{GHz} \) is observed at \( T = 35 \, \text{mK} \), despite the low frequency resolution. This result is obtained with a sample size of \( 5.5 \, \text{mm} \times 30 \, \mu\text{m} \). Notably, 
the same saturation frequency \( f_s \sim 2.4 \, \text{GHz} \) is 
consistently observed across Hall bars of different sizes.Ref.\cite{exp8} }
\label{101f}
\end{figure}

\vspace{0.1cm}
Furthermore, it is clear that the saturation frequency does not depend on the size of the Hall bar. The experiment shown in Fig.(\ref{101f}) demonstrates that a similar saturation frequency of \( 2.2 \, \text{GHz} \sim 2.5 \, \text{GHz} \) is observed in a smaller sample measuring \( 5.5 \, \text{mm} \times 30 \, \mu m \), with a surface area of \( 1.6 \times 10^{-3} \, \text{cm}^2 \). The material (\( \rm GaAs/AlGaAs \)) and experimental setup are identical to those used in the larger sample, although the frequency resolution is slightly lower. This independence of the saturation frequency \( f_s = m_a/2\pi \) from the size of the Hall bar is a crucial feature of the axion effect. (A similar discussion regarding the approximate independence of saturation temperature on the size of the Hall bar when the axion effect is present can be found in Fig.(\ref{6f}).)

\begin{figure}[htp]
\centering
\includegraphics[width=0.95\hsize]{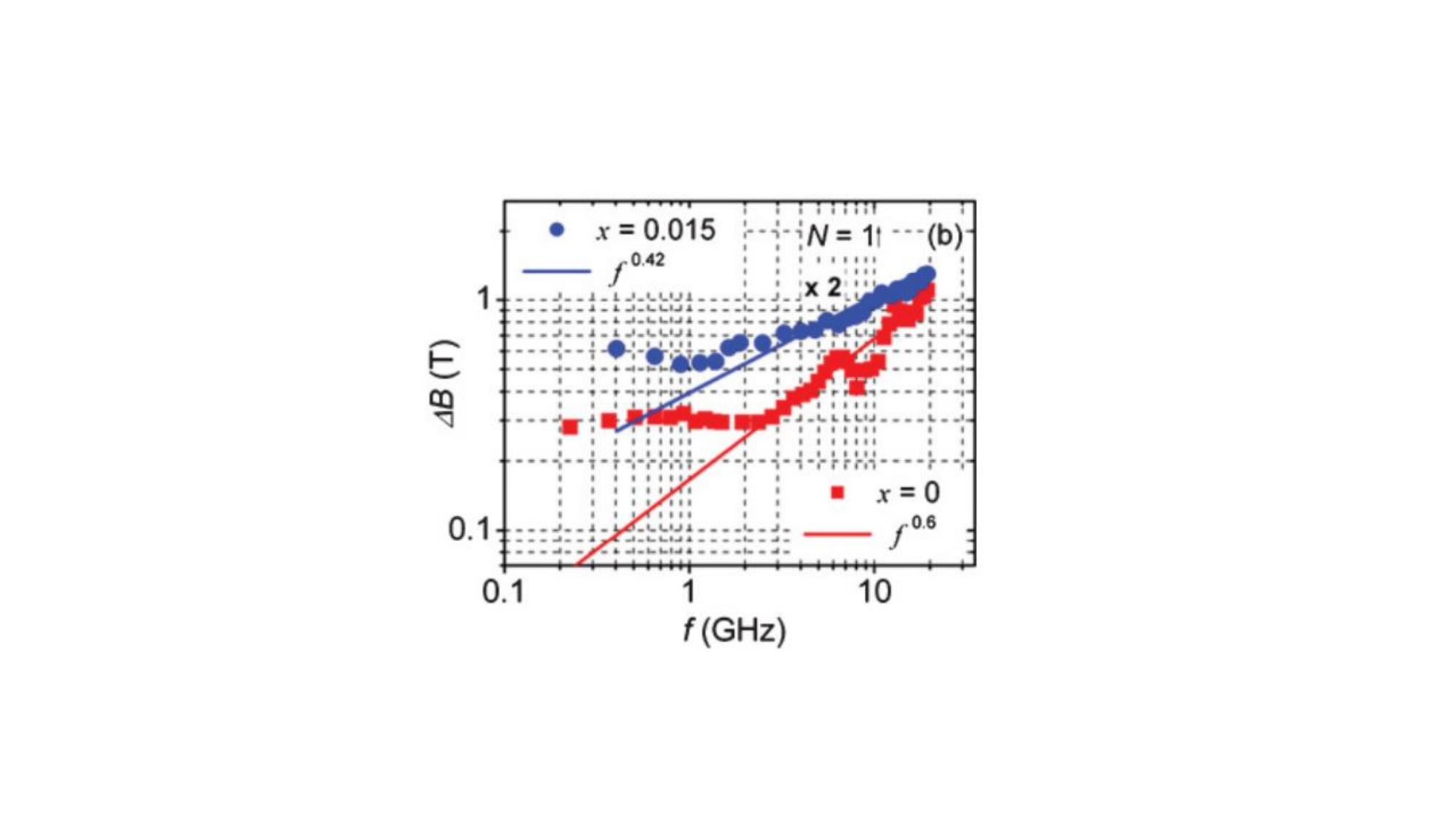}
\caption{The lower curve shows a similar saturation frequency of \( f_s \sim 2.4 \, \text{GHz} \) for a different plateau transition labeled \( N1 \uparrow \), distinct from the transition labeled \( N1 \downarrow \) in Fig. \ref{10f}, using the same sample at a temperature of \( 35 \, \text{mK} \). The upper curve represents results obtained from a sample of \( \text{Al}_{0.015}\text{Ga}_{0.985}\text{As}/\text{Al}_{0.32}\text{Ga}_{0.68}\text{As} \), which contains a small amount of Al contamination. Ref.(\cite{exp8})}
\label{110f}
\end{figure}

\vspace{0.1cm}
We present additional evidence supporting the presence of the axion. The experiment shown in Fig.(\ref{110f}) reveals that the saturation frequency is identical across different plateau-to-plateau transitions, as indicated by the transitions labeled \( N \downarrow \) 
in Fig.(\ref{10f}) and Fig.(\ref{101f}), and \( N \uparrow \) in Fig.(\ref{110f}). This holds true despite the lower frequency resolution in these measurements.

In the absence of the axion effect, different plateau-to-plateau transitions would result in distinct saturation frequencies, as demonstrated in Fig.(\ref{9f}). This occurs because the mobility gap \( \delta(T, f) \) varies with each plateau-to-plateau transition. The saturation of the mobility gap at \( f = f_c \) leads to the saturation of \( \Delta E_f \), with \( f_s = f_c \) because of
$\Delta E_f\sim 2\delta (f)$.
However, when the axion effect is present, we observe a uniform saturation frequency, \( f_s = m_a/2\pi \), regardless of the plateau-to-plateau transition or the size of the Hall bar.


\vspace{0.1cm}
We have discussed how the saturation frequency remains independent of the size of the Hall bar and the specific plateau-to-plateau transition as long as the axion effect is significant. Based on the above arguments, we expect the following behavior of 
saturation frequency.
As we gradually increase the temperature in the above experiment, making the axion effect less pronounced, the saturation frequency begins to depend on temperature and decreases to a value below 1 GHz, as expected in the absence of the axion effect. However, beyond a certain point, further increases in temperature lead to a rise in the saturation frequency, as shown in Fig. (\ref{7f}). The expectation 
comes from the fact that the reduction of the axion effect results in a decrease in the saturation frequency from 2.4 GHz.

\vspace{0.1cm}
A similar effect occurs when we increase the impurity concentration of aluminum (\(x\)) in the sample \(\rm Al_xGa_{1-x}As/Al_{0.32}Ga_{0.68}As\). The sample in Fig.(\ref{10f}) has \(x = 0\), and as \(x\) increases, the axion effect gradually diminishes. This results in a gradual decrease in the saturation frequency \(f_s\) from 2.4 GHz. However, once \(x\) surpasses a critical value, the saturation frequency stops decreasing and starts to increase.

In fact, as seen in Fig.(\ref{110f}), a slightly lower saturation frequency of \(f_s(35 \, \text{mK}) \sim 1.5\) GHz was obtained using a sample of \(\rm Al_xGa_{1-x}As/Al_{0.32}Ga_{0.68}As\) with small contamination (\(x = 0.015\)) and a large size (\(22 \, \text{mm} \times 30 \, \mu \text{m}\)). This sample includes both short-range disorder potential \(V_s\) and long-range disorder potential \(V_l\), whereas the sample \(\rm GaAs/AlGaAs\) only includes \(V_l\). The small contamination of \(V_s\) leads to a larger mobility gap \(\Delta E\), which reduces the axion effect as described in equation (\ref{17}). The result is consistent with our expectations: the small contamination of \(x = 0.015\) causes the saturation frequency to decrease from \(f_s = 2.4\) GHz to \(f_s = 1.5\) GHz.

In contrast, the previous experiment shown in Fig.(\ref{5f}) used a sample with high aluminum contamination (\(x = 0.85\)), where the axion effect is negligible. As anticipated, the experiment demonstrates a formula for the saturation frequency \(f_s(W) \propto 1/W\), where the saturation frequency depends on the size \(W\) of the Hall bar. This behavior further confirms the absence of the axion effect in this sample.

\vspace{0.1cm}
Although we previously mentioned that the frequency \( f \) in the formula \(\Delta E_f = 2\delta(T, f) + 8\pi f\) is not the real frequency, the saturation frequency \( f_s = m_a/2\pi \) represents the actual frequency of the axion microwave. This is because we are comparing the measured microwave frequency with the frequency of the axion microwave. Since both are microwaves, they smooth out the electron distribution in the same manner. Therefore, the relation \( f_s = m_a/2\pi \) directly corresponds to the real axion mass.

\vspace{0.1cm}
The experiments with large Hall bars, as mentioned above, use a coplanar waveguide to impose microwaves. This setup differs from those using smaller Hall bars. Hence, a straightforward comparison of the results should be made with caution. In the following, we present another experiment\cite{exp10} that also utilizes a coplanar waveguide.

%
%
%
%
%
%
%
%

\begin{figure}[htp]
\centering
\includegraphics[width=0.7\hsize]{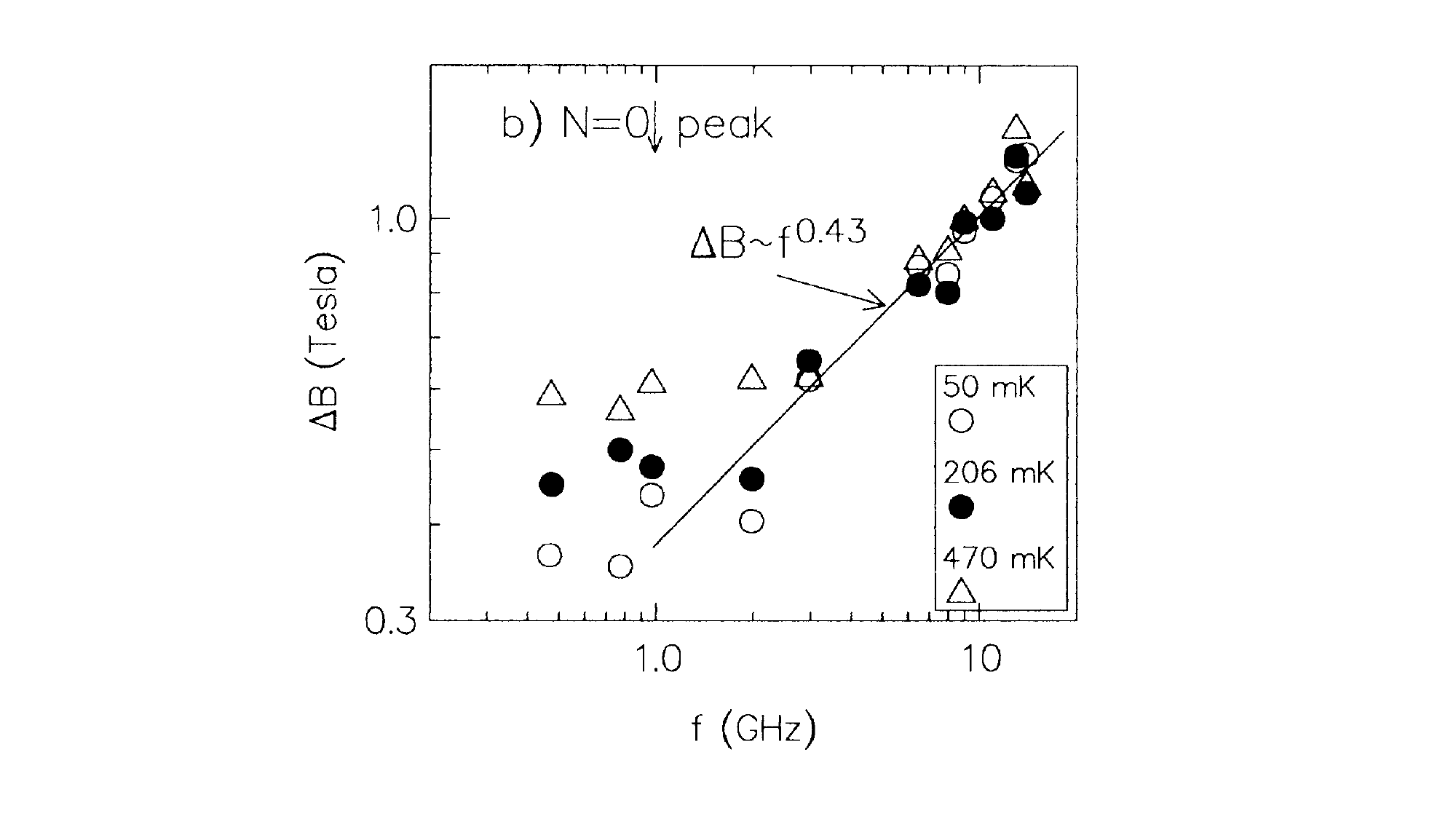}
\caption{A saturation frequency of \( f_s \sim 0.8 \, \text{GHz} \) is observed at \( 50 \, \text{mK} \) in a relatively large sample measuring approximately \( 14 \, \text{mm} \times 30 \, \mu\text{m} \). However, this sample likely contains a high level of aluminum contamination, specifically in the form of \( \text{Al}_x\text{Ga}_{1-x}\text{As}/\text{GaAs} \) with a significantly higher aluminum fraction (\( x \)) than \( x = 0.015 \). This inference is based on the observed low electron mobility of \( \mu = 4 \times 10^4 \, \text{cm}^2/\text{Vs} \).
Ref.\cite{exp10} }
\label{11f}
\end{figure}

The results are shown in Fig.(\ref{11f}) for a large Hall bar with dimensions \(14\ \text{mm} \times 30\ \mu\text{m} \sim 4 \times 10^{-3} \ \text{cm}^2\), and temperatures of \(470\ \text{mK}\), \(206\ \text{mK}\), and \(50\ \text{mK}\), although the frequency resolution is quite low. The sample is composed of \(\rm Al_xGa_{1-x}As/GaAs\) with a much larger Al concentration than \(x = 0.015\). The mobility is relatively low, with \(\mu = 4 \times 10^4 \ \text{cm}^2/\text{Vs}\).

The saturation frequencies observed are approximately,
\(f_s(470\ \text{mK}) \sim 3\ \text{GHz}\), 
\(f_s(206\ \text{mK}) \sim 2\ \text{GHz}\) and 
\(f_s(50\ \text{mK}) \sim 0.8\ \text{GHz}\)

These results are consistent with those where no axion effect is observed. Despite the low temperature (\(T = 50\ \text{mK}\)) and large sample size, the axion effect does not appear, likely due to the large \(\Delta E_s\) induced by the higher Al contamination or the reduced mobility. This contrasts with the experiment shown in Fig.(\ref{10f}), where the sample with no Al contamination (\(x = 0\)) and larger mobility (\(\mu = 3.8 \times 10^5 \ \text{cm}^2/\text{Vs}\)) shows observable axion effects.

\vspace{0.1cm} 
The results also show that the width \(\Delta B\) remains independent of temperature for \(f > 3\ \text{GHz}\), implying that \(\Delta E_f = 2\delta(T, f) + 8\pi f \sim 8\pi f\) for \(f > 3\ \text{GHz}\). 
However, for frequencies below \(3\ \text{GHz}\), the mobility gap \(\delta(T, f)\) becomes comparable to the frequency \(f\).
(In the smaller Hall bar experiment shown in Fig.(\ref{7f}), the width \(\Delta B\) at different temperatures coincides for frequencies greater than 8 GHz, which is attributed to the larger mobility gap in the smaller Hall bar.) 

\vspace{0.2cm}
We have explored the dependence of saturation frequency on temperature, sample size, and material. When the sample does not experience the axion effect, the saturation frequency \(f_s\) generally decreases with decreasing temperature or increasing Hall bar size. For instance, at low temperatures (e.g., \(150\ \text{mK}\)) and in small samples (e.g., \(164\ \mu\text{m} \times 64 \mu\text{m} \sim 10^{-4} \ \text{cm}^2\)), the saturation frequency is typically less than 1 GHz. However, experiments\cite{exp9, exp8} with a large sample at a low temperature of \(35\ \text{mK}\) show a saturation frequency of around \(2.4\ \text{GHz}\). This frequency is independent of Hall bar size and plateau-plateau transitions, similar to the observation that saturation temperature \(T_s = m_a\) is independent of Hall bar size. This result is unexpected and suggests that the saturation may be due to the axion effect.

We also demonstrate that by slightly increasing the Al contamination, which diminishes the axion effect, the saturation frequency decreases from \(2.4\ \text{GHz}\). These experiments provide strong evidence for the presence of the axion, with its mass \(m_a = 2\pi f_s\), yielding \(m_a \sim 10^{-5}\ \text{eV}\).

\section{several ways of confirmation of axion contribution}

In the experiment shown in Fig.(\ref{10f}) from \cite{exp9}, decreasing the external frequency reveals that the width \(\Delta B\) saturates at a high frequency, around \(2.4\ \text{GHz}\). This high saturation frequency has been discussed as an indication of the axion microwave effect. Without the axion contribution, the saturation frequency would be expected to occur at a much lower value, below \(1\ \text{GHz}\).

To test if the observed saturation is indeed due to the axion, we propose shielding the axion-generated radiation and observing whether the saturation frequency subsequently decreases below \(1\ \text{GHz}\). In our previous work \cite{iwa}, we suggested using two parallel conducting plates positioned parallel to the magnetic field, sandwiching the Hall bar to block axion-generated radiation from external sources. Although these plates generate microwaves, they are not absorbed by the Hall bar because their electric field component is oriented perpendicular to the Hall bar's surface.

Additionally, conducting plates placed perpendicularly to the magnetic field \(\vec{B}\) do not generate additional microwaves in the presence of the axion-induced oscillating electric field \( \vec{E}_a \propto \cos(m_a t) \vec{B} \). 
Instead, axion-generated microwaves originate from oscillating electric currents induced by $\vec{E}_a$
in nearby metals that are not necessarily aligned perpendicular to \( \vec{B} \). These metals are typically found surrounding the Hall bar in quantum Hall effect experiments.

\begin{figure}[htp]
\centering
\includegraphics[width=0.9\hsize]{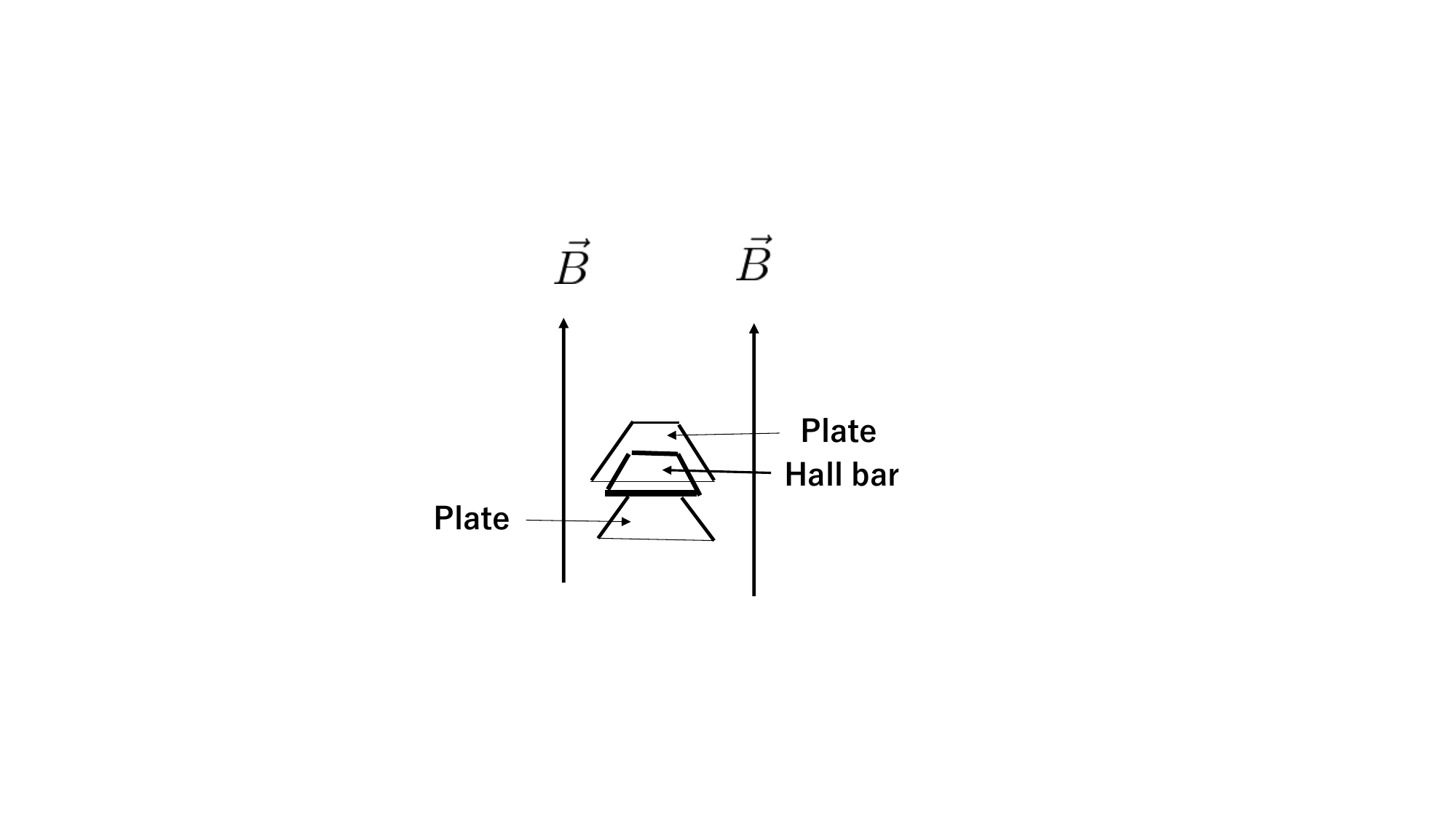}
\caption{Hall bar sandwiched by two conducting plates parallel to the Hall bar, which shield axion microwave}
\label{12f}
\end{figure}

\vspace{0.1cm}
In this paper, we propose an alternative setup: positioning the shielding plates parallel to the Hall bar, as shown in Fig.(\ref{12f}). These plates would be oriented perpendicularly to \(\vec{B}\), ensuring they do not produce axion-induced radiation. To test this configuration, we suggest conducting measurements using two identical Hall bars, one with shielding plates and one without. Both Hall bars would have sufficiently large dimensions and be maintained at low temperatures to allow the axion effect to manifest.

Comparing the saturation frequencies in each setup, we would expect to observe a high frequency (around \(2.4\ \text{GHz}\)) in the unshielded Hall bar and a lower frequency (below \(1\ \text{GHz}\)) in the shielded Hall bar if the axion effect is present. Finding such a discrepancy in \(f_s\) between the shielded and unshielded configurations would provide evidence for the presence of axion dark matter.

 \vspace{0.1cm}
We would like to propose another ways of the confirmation of the axion effect in integer quantum Hall effect.

Firstly, 
we measure the dependence of saturation frequency $f_s(T)$ on temperature.
If we find that the saturation frequency $f_s(T)$ is given by $2.4$GHz at low temperature like $30$mK,
we check whether $f_s(T)$ varies as decreasing temperature. When the axion effect causes $f_s(T=30\rm mK)= 2.5$GHz,
the saturation frequency does not vary even if we decrease the temperature, i.e. $f_s(T\le 30\rm mK)=2.5$GHz.
After that, we increase temperature larger than $30$mK and check whether $f_s(T)$ decreases or not.
We expect that $f_s(T)$ begins to decrease
when the temperature becomes sufficiently large for the axion effect to be negligible. 
There is a critical temperature above which the frequency $f_s(T)$ begins to decrease.
The axion effect gradually disappears as the temperature increases beyond the critical one. Then,  
the frequency  $f_s(T)$ reaches a low frequency $<1$GHz at a temperature $T_c$.
When the temperature goes beyond $T_c$, the frequency $f_s(T)$ begins to increases just as
in the case without the axion effect. It is our expectation when the dark matter axion is present.

\vspace{0.1cm}
Secondly, we can check the saturation frequency $\sim 2.4$GHz
does not vary with increasing the size of Hall bar.
When we observe the saturation frequency $\sim 2.4$GHz with a large Hall bar at sufficiently low temperature,
the frequency does not change even if the size of Hall bar is enlarged. 
The independence of $f_s\sim 2.4$GHz on the size of Hall bar is caused by the axion effect.

\vspace{0.1cm}
Finally, we propose an additional check of the confirmation. We examine whether or not 
the identical saturation frequency $f_s\sim 2.4$GHz is obtained even if we use 
Hall bars formed of various components of semiconductors. But it is favorable that they do not involve 
short range disorder potential, which induces large extension $\Delta E$ in density of state $\rho(E)$.
The examination should be performed 
in sufficiently large size and low temperature for
the axion effect to be effective. The saturation frequency is determined by the frequency of the axion
microwave, not depending on each samples. 

\vspace{0.1cm}
Theses examinations using various ways of the confirmation will be able to prove the presence of the axion dark matter.

\section{conclusion}

By analyzing previous experiments, we have shown how the axion effect manifests in plateau-plateau transitions. These transitions are characterized by the width $\Delta B$, 
which depends on temperature, the frequency of the external microwave, and the size of the Hall bar. 
The transition width $\Delta B$ is related to the change in Fermi energy, $\Delta E_f$, which increases as the magnetic field $B$ decreases. Specifically, the width $\Delta E_f$ is given by $\Delta E_f=2\delta +4T$
in the temperature-dominant case, where $\delta $ is the mobility gap, 
or $\Delta E_f=2\delta +8\pi f$ in the frequency-dominant case, with $f$ being the microwave frequency.


\vspace{0.1cm}

In general, the width $\Delta B$ decreases with decreasing temperature, but it saturates at a critical temperature $T_c$. In the case of a small Hall bar, the mobility gap $\delta(T)$ is much larger than the thermal energy, satisfying $2\delta(T) \gg 4T$. Thus, the width saturation occurs when $\Delta E_f(T) = 2\delta(T) + 4T \approx 2\delta(T)$, stabilizing at $T = T_c$.

\vspace{0.1cm}
As the mobility gap $\delta(T)$ decreases with the increasing size of the Hall bar, the critical temperature $T_c$ also decreases with larger Hall bar dimensions. This scenario corresponds to the axion effect being negligible. In contrast, at much lower temperatures and for larger Hall bar sizes, the axion effect becomes significant.


Specifically, in the case of a sufficiently large Hall bar, where the axion effect cannot be neglected, the width $\Delta B$ saturates at a temperature $T_s = m_a$, with $\Delta E = 2\delta(T_s) + 4m_a$ in our formalism. An experiment \cite{exp4} that observed this saturation showed a temperature below $30$mK and a two-dimensional electron surface area exceeding $10^{-3} \ \rm{cm}^2$. In this setup, the mobility gap is comparable to or smaller than the thermal energy. Notably, the saturation temperature $T_s = m_a$ almost remains constant regardless of Hall bar size. 

We have shown in section(\ref{axion}) that actual saturation temperature $T_s$ below which the axion effect dominates thermal effect,
is, roughly speaking, a ten times smaller than $m_a$. So, it is not precise relation $T_s=m_a$.
Furthermore, it slightly depends on parameters of the samples.

\vspace{0.1cm}
The approximate size independence of the saturation temperature, demonstrated experimentally in \cite{exp4}, has been attributed to "intrinsic decoherence," although it is a prominent characteristic of the axion effect in the quantum Hall regime. The saturation of $\Delta B$ is illustrated in Fig.(\ref{6f}) of the reference\cite{exp4}.

\vspace{0.1cm}
In fact, an experiment \cite{exp44} conducted with a different sample from that used in \cite{exp4} demonstrates that the width $\Delta B$ saturates at nearly the same temperature as observed in \cite{exp4}.



\vspace{0.1cm}

Similarly, the width $\Delta B$ decreases as the frequency $f$ of the external microwave decreases, eventually saturating at a frequency $f_s(T)$. This saturation frequency decreases with temperature, typically falling below $1$ GHz 
at around $100$mK. For example, $f_s \approx 0.8$ GHz when the temperature is $150$ mK 
and the Hall bar size is $164\rm{\mu m} \times 64 \ \rm{\mu m}$. As the Hall bar size increases (or temperature decreases) beyond this point, $f_s(T)$ becomes significantly lower, as far as it is a regime where the axion effect is negligible.

Experiments \cite{exp5,exp6,hohls2} exhibiting these characteristics are shown in Fig.(\ref{7f}), Fig.(\ref{8f}), and Fig.(\ref{9f}). Additionally, the authors of \cite{hohls2} reported that in various experiments conducted up to the publication date, the saturation frequencies were consistently below $1$GHz at approximately $100$mK.

\vspace{0.1cm}

Conversely, when the axion effect is not negligible, the saturation frequency is precisely given by $f_s = m_a / 2\pi$.
( The relation is accurate, while the relation of the saturation temperature $T_s=m_a$ derived from the assumption of cut off in electron distribution is not. ) 
This condition is realized at low temperatures below $100$mK and with a large Hall bar area exceeding $10^{-3} \rm{cm}^2$. In the absence of the axion effect, the saturation frequency would be expected to be below $1$GHz due to the low temperature and large Hall bar size. However, we observed $f_s \approx 2.4$ GHz in experiments \cite{exp8,exp9} with a large Hall bar ($22 \rm mm\times 30 \mu$m) at a low temperature of $35$mK.

As expected, the saturation frequency remains unchanged \cite{exp9} even with variations in Hall bar size. Additionally, an identical saturation frequency \cite{exp8} was observed during a plateau-to-plateau transition with a different filling factor. Consequently, we conclude that the experiments \cite{exp8,exp9}, shown in Fig. (\ref{10f}), provide strong evidence for the presence of axion dark matter with a mass of approximately $m_a \approx 10^{-5} \ \rm{eV}$.



\vspace{0.1cm}

To confirm the presence of the axion effect, we propose experiments to test whether the saturation frequency $f_s = m_a / 2\pi \approx 2.4$ GHz remains constant even when varying the temperature, Hall bar size, and filling factor in a plateau-to-plateau transition.

A more direct method for confirmation would involve shielding the axion-induced microwave signal as illustrated in Fig. (\ref{12f}) to determine if the phenomena associated with the axion effect disappear under these conditions.

%

\vspace{0.2cm}
The author
expresses thanks to A. Sawada and Wen Yin for useful comments.
This work is supported in part by Grant-in-Aid for Scientific Research ( KAKENHI ), No.19K03832.



\end{document}